\documentclass[aps,pre,twocolumn,floatfix]{revtex4}

\usepackage{graphicx}
\usepackage{amssymb,amsfonts,amsmath}
\usepackage{epsf}
\usepackage{subfigure}
\usepackage{epstopdf}

\newcommand{\be}{\begin{equation}}
\newcommand{\ee}{\end{equation}}
\newcommand{\bea}{\begin{eqnarray}}
\newcommand{\eea}{\end{eqnarray}}

\begin{document}

\title{\bf Kinetics and thermodynamics of exonuclease-deficient DNA polymerases}

\author{Pierre Gaspard}
\affiliation{Center for Nonlinear Phenomena and Complex Systems,\\
Universit\'e Libre de Bruxelles, Code Postal 231, Campus Plaine,
B-1050 Brussels, Belgium}

\begin{abstract}
A kinetic theory is developed for exonuclease-deficient DNA polymerases, based on the experimental observation that the rates depend not only on the newly incorporated nucleotide, but also on the previous one, leading to the growth of Markovian DNA sequences from a Bernoullian template.  The dependences on nucleotide concentrations and template sequence are explicitly taken into account.  In this framework, the kinetic and thermodynamic properties of DNA replication, in particular, the mean growth velocity, the error probability, and the entropy production in terms of the rate constants and the concentrations are calculated analytically.  Theory is compared with numerical simulations for the DNA polymerases of T7 viruses and human mitochondria.
\end{abstract}

\noindent 
\vskip 0.5 cm

\maketitle

\section{Introduction}

Biological systems are characterized by their metabolism and self-replication.  If the former refers to the consumption and dissipation of energy maintaining the system out of thermodynamic equilibrium, the latter requires the faithful transmission of genetic information between successive generations.  Since self-replication is powered by the metabolism, a fundamental coupling exists between energetic and genetic aspects in living organisms.

At the molecular level, genetic information is coded in DNA sequences of nucleotides (nt).  These macromolecules are aperiodic copolymers composed of four types of monomeric units $\{{\rm A},{\rm C},{\rm G},{\rm T}\}$.  During replication, information is transmitted by the copolymerization of a new DNA strand along the template formed by an old DNA strand.  The synthesis is catalyzed by an enzyme called DNA polymerase and powered by the chemical energy of about two adenosine triphosphates per incorporated nucleotide \cite{Alberts,LBSK58,BLSK58,BK72}. 

Thermal fluctuations are ambient at the molecular scale so that errors may occur during the replication process, possibly causing mutations.  These errors generate some disorder in the growing sequence.  Remarkably, the thermodynamic entropy production of copolymerization depends on this disorder, establishing a fundamental link between thermodynamics and molecular information processing \cite{B79,AG08,ELV10,WW11,G15}.  This link is in action during DNA replication.  Indeed, DNA polymerases can be very efficient in transmitting genetic information with surprisingly low error probability as small as $10^{-5}$-$10^{-6}$, even without dedicated proofreading mechanisms such as the exonuclease activity or the postreplication DNA mismatch repair \cite{Alberts,L74,SR83,IPBM06}.  Such low error probability cannot be explained in terms of free energy for base pairing.  In fact, the difference of free energy between correct (Watson-Crick) and incorrect base pairs is about $\Delta\Delta G_{\rm bind}\simeq  14$~kJ/mol, corresponding to an error probability of the order of $10^{-2}$ \cite{LK82}.  In the seventies, Hopfield and Ninio showed that kinetics can amplify the discrimination between correct and incorrect pairings, very much reducing the error probability when replication is driven out of equilibrium \cite{H74,N75}.  The biochemistry of DNA polymerases has been systematically investigated, providing experimental data on the rate constants for the formation of the sixteen possible base pairs at the growing end of DNA \cite{LK82,PWJ91,WPJ91,DPJ91,J93,TJ06,JJ01a,JJ01b,LNKC01,LJ06,EG91,KB00,SUKOOBWFWMG08,RBT08,ZBNS09,DJPW10,BBT12}.  Furthermore, these studies have revealed that DNA polymerases undergo conformational changes during DNA elongation and that the copolymerization process depends not only on the new nucleotide that is attached to the growing copy, but also on the previously incorporated nucleotide, providing the polymerares with molecular mechanisms to detect mismatches and react accordingly \cite{J93}.

Till now, the challenges for theoretical modelling such a dependence have prevented the development of kinetic theory describing these essential aspects of DNA polymerases.  In early theoretical studies \cite{B79,AG08,CQ09,SC12,SP13,S14,MHL12,MHL14}, the kinetic schemes have supposed that the rates depend only on the newly incorporated nucleotide, although the effect of possible correlations between consecutive steps has already been envisaged for some limiting cases \cite{RP15}.  Recently, theoretical methods have been developed to determine rigorously the properties of molecular chains growing with attachment and detachment rates also depending on the previously incorporated nucleotide \cite{GA14}.  In this context, the inclusion of detachment events is {\it sine qua non} to obtain finite thermodynamic quantities and, in particular, the entropy production \cite{AG08,CQ09}.  These issues about kinetics also concern DNA transcription by RNA polymerases and the translation to proteins by ribosomes \cite{JB98,WEMO98,W06,LFW07,VCML09,DPG13,GCCR09}.

In the present and companion~\cite{paperII} papers, our purpose is to develop a kinetic theory of DNA polymerases taking into account the effects of the previously incorporated nucleotide, which is a crucial aspect of these enzymes.

The present paper is focused on exonuclease-deficient DNA polymerases.  These enzymes are obtained by mutagenesis in order to study the copying fidelity in the absence of exonuclease proofreading and their kinetic properties are measured {\it in vitro} with the experimental techniques of biochemistry \cite{PWJ91,WPJ91,TJ06,JJ01a,LNKC01,LJ06}.  The kinetic equations and thermodynamics will be presented in Section~\ref{Kin+Thermo} and Appendix~\ref{AppA}.  The theory is set up to include the dependence on the concentrations of nucleotides and other substances, which are the direct control parameters of biochemical processes such as DNA replication \cite{ZQQ12,GQQ12,C13}.  The theory also emphasizes the dependence on the copy and template sequences.  Indeed, the template constitutes a disordered medium for the random drift of the enzyme at the growing end of the copy \cite{JB98,WEMO98}.  By including these different dependences, the theory is suited for dealing with experimental data from biochemistry.  In particular, the adopted kinetic scheme reproduces the Michaelis-Menten kinetics of DNA polymerases with its characteristic dependence on nucleotide concentrations~\cite{MM13,JG11}, as it should to compare with experiments.  In Section~\ref{B-chain}, the kinetic equations are solved analytically in the simple case where the rates only depend on correct or incorrect pairing of the newly incorporated nucleotide, leading to the growth of a Bernoulli chain. In Section~\ref{M-chain} and Appendix~\ref{AppB}, analytical methods are given if the rates also depend on the previously incorporated nucleotide, which generates instead a Markov chain.  These methods are applied to the DNA polymerases of T7 viruses and human mitochondria in Sections~\ref{T7-Pol} and~\ref{Hum-Pol}.  The algorithm used for numerical simulations is described in Appendix~\ref{AppC}.  A discussion is carried out in Section~\ref{Discussion}.

The companion paper will be devoted to DNA polymerase with exonuclease proofreading, in which case the dependence of the rates on the previously incorporated nucleotide will turn out to be essential~\cite{paperII}.

\section{Kinetics and thermodynamics}
\label{Kin+Thermo}

\subsection{Generalities}

DNA polymerases are enzymes catalyzing the synthesis of DNA from
the four deoxyribonucleoside triphosphates dATP, dCTP, dGTP, and dTTP, 
more shortly, the nucleotides:
\be
{\rm dNTP} \ + \ {\rm E}\cdot{\rm DNA}_l
  \quad \rightleftharpoons
  \quad
{\rm E}\cdot{\rm DNA}_{l+1} \ + \ {\rm PP}_{\rm i} \; .
\label{pol-react}
\ee
Pyrophosphate PP$_{\rm i}$ is released following the incorporation of nucleotides and the elongation of the DNA polymer.  The copolymerization proceeds along a template made of a single-stranded DNA (ssDNA), leading to DNA replication.  This nonequilibrium process is powered by the metabolism
with the chemical free energy of about two adenosine triphosphates per incorporated nucleotide.

DNA polymerases consist of a complex of several proteins.  The domains of polymerase and exonuclease activities can be found either on the same polypeptide (e.g. for pol. I, T4 DNA pol., T7 DNA pol., human mitochondrial DNA pol. $\gamma$), or on separate polypeptides (e.g. for pol. III) \cite{J93}.  The exonuclease activity can be essentially switched off by mutagenesis, yielding exonuclease-deficient (exo$^-$) mutants.  The purpose of the present paper is to set up a minimal kinetic theory of exonuclease-deficient DNA polymerases, explicitly establishing the dependence of copolymerization on the concentrations of the different possible substances (dATP, dCTP, dGTP, dTTP, and PP$_{\rm i}$), and the template and copy sequences.  This framework allows us to obtain the thermodynamic quantities and to deduce analytic expressions for the error probability in terms of the concentrations and the reaction constants for the different regimes close and away from equilibrium.  

The overall reaction~(\ref{pol-react}) summarizing the polymerase activity is composed of several elementary steps that have been analyzed by Johnson and coworkers \cite{PWJ91,WPJ91,DPJ91,J93,TJ06,JJ01a,JJ01b,LNKC01,LJ06}, as well as other groups \cite{LK82,EG91,KB00,SUKOOBWFWMG08,RBT08,ZBNS09,DJPW10,BBT12}.  The rate-limiting steps are conformational changes of the enzyme, playing an essential role in the processive nucleotide incorporation \cite{J93}.  The two main steps of the polymerase activity are: (1)~the binding of dNTP to the template with the formation of a correct Watson-Crick base pair or an incorrect one; (2)~the release of pyrophosphate PP$_{\rm i}$ and the incorporation of dNMP by the formation of a phosphodiester bond between the dNMP and the growing DNA chain.  In order to study thermodynamics, we need to include the reverse reactions that are: (1)~the dissociation of dNTP; (2)~the pyrophosphorolysis of the nucleotide at the end of the copy by a PP$_{\rm i}$ molecule coming from the surrounding aqueous solution.

Because of molecular and thermal fluctuations, each step may randomly occur at rates given by the kinetics.  The key point is that DNA polymerization is controlled by the concentrations of nucleotides dNTP and pyrophosphate PP$_{\rm i}$ in the surrounding solution.  This latter is supposed to be large enough to act as an infinite reservoir so that the concentrations of dNTP and PP$_{\rm i}$ are kept constant during the process.  Consequently, the chemical potentials of these species also remain constant in time:
\be
\mu_{\rm X} = \mu_{\rm X}^0 + RT \ln \frac{[{\rm X}]}{c^0} \, ,
\ee
where X = dATP, dCTP, dGTP, dTTP, or PP$_{\rm i}$; $R$ is the molar gas constant; $T$ is the temperature; $[{\rm X}]$ denotes the concentration of X; and $c^0=1$\,M is the standard reference concentration.

Copolymerization proceeds if the dNTP concentrations exceed a threshold proportional to the PP$_{\rm i}$ concentration, otherwise the DNA copy may undergo depolymerization.  For exonuclease-deficient DNA polymerases, thermodynamic equilibrium happens at a threshold concentration where the growth velocity of the copy is vanishing.  Under normal physiological conditions, the concentrations of dNTP and PP$_{\rm i}$ take the following values \cite{T94,H01}:
\bea
&&[{\rm dNTP}] \simeq \mbox{5-40 $\mu$M} \, , \\
&&[{\rm PP}_{\rm i}] \simeq \mbox{0.2-0.3 mM} \, .
\eea
These values may vary between the nucleus and the cytoplasm and during the cell cycle. Imbalances of the intracellular dNTP pool may be linked to cancer, genetic diseases, and biological mutagenesis \cite{VGNV05}.

Template-directed copolymerization also depends on the sequence of the template.  Under the aforementioned conditions, the motion of the enzyme along the template is a biased diffusion with a mean drift velocity powered by the chemical free energy of the reaction~(\ref{pol-react}).  This biased diffusion process takes place along the aperiodic chain of the template.  On this disordered medium, copolymerization may thus undergo stochastic switches between forward and backward movements depending on the random occurrence of subsequences favorable or unfavorable to the growth.

As emphasized in the introduction, the kinetics of DNA polymerases is highly sensitive to the nucleotide previously incorporated in the growing copy, allowing an important discrimination between correct and incorrect pairings.  Therefore, our minimal theory should take into account the sequences of both the copy and the template.

We notice that copolymerization may be interrupted by the dissociation of the enzyme from DNA:
\be
{\rm E}\cdot{\rm DNA}_l
  \quad \overset{k_{\rm off}}{\underset{k_{\rm on}}{\rightleftharpoons}}
  \quad
{\rm E} \ + \ {\rm DNA}_l \, .
\label{E+DNA}
\ee
The dissociation rate $k_{\rm off}$ combined with the maximal polymerization rate $k^{\rm p}_{+,{\rm max}}$ gives an estimation of the so-called processivity \cite{J93}, i.e., the maximal number of nucleotides incorporated before an interruption, $l_{\rm max}\simeq k^{\rm p}_{+,{\rm max}}/k_{\rm off}$, which is often large enough to justify that the dissociation~(\ref{E+DNA}) is neglected.

\subsection{Kinetic scheme}

\begin{figure}[h]
\centerline{\scalebox{0.7}{\includegraphics{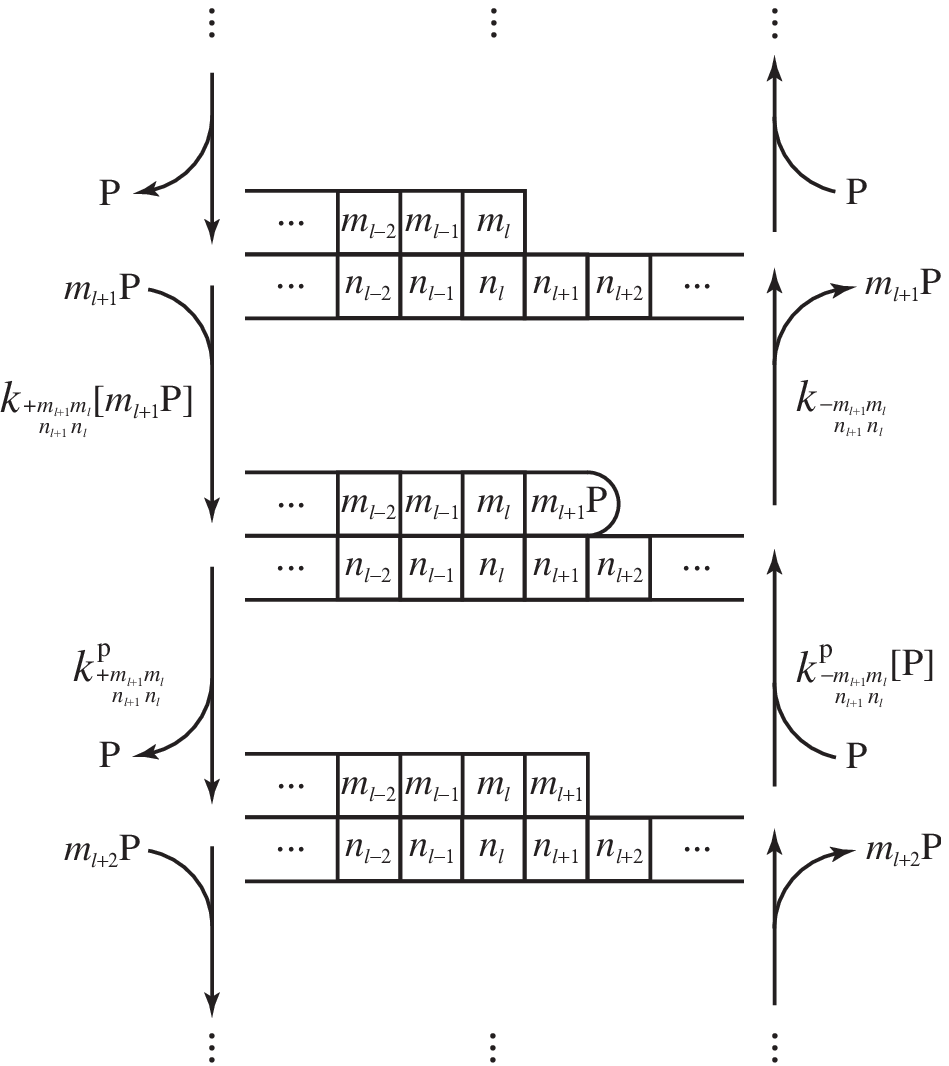}}}
\caption{Kinetic scheme of the polymerase activity.  $\{m_j\}$ denotes the ssDNA copy, $\{n_j\}$ the ssDNA template, $m_j{\rm P}$ deoxynucleoside triphosphates dNTP, and P pyrophosphates PP$_{\rm i}$.}
\label{fig1}
\end{figure}

The kinetics of DNA polymerases is explicitly formulated in terms of the sequences of nucleotides in the template and the copy, which provides a complete description of the process.
Figure~\ref{fig1} depicts the simplified kinetic scheme we here consider for exonuclease-deficient polymerases.  The mass action law determines the reaction rates of the elementary steps.  The sequences of the template $\alpha=n_1\cdots n_l n_{l+1}\cdots$ and of the copy $\omega=m_1\cdots m_l$ are composed of successive nucleotides $m,n\in\{ {\rm A}, {\rm C}, {\rm G}, {\rm T}\}$.  An essential aspect of the kinetics is that the rates depend not only on the nucleotide $n_l$ of the template because of the formation of the base pair $m_l$:$n_l$, but also on the previously incorporated nucleotide $m_{l-1}$ and its correct or incorrect pairing $m_{l-1}$:$n_{l-1}$.

Starting from a copy of length $l$ with the ultimate monomeric unit $m_l$, the next forward reaction is the binding of the deoxyribonucleoside triphosphate $m_{l+1}{\rm P}$ at the
\be
\mbox{nucleotide binding rate:} \qquad k_{+m_{l+1} m_l\atop \ \, n_{l+1} \, n_l}[m_{l+1}{\rm P}] \, ,
\label{nb_rate}
\ee
which is proportional to the concentration $[m_{l+1}{\rm P}]$ of this nucleotide in the surrounding solution.
Thereafter, the pyrophosphate PP$_{\rm i}$ -- denoted P -- is released at the
\be
\mbox{polymerization rate:} \qquad k^{\rm p}_{+m_{l+1} m_l\atop \ \, n_{l+1} \, n_l} \, ,
\label{pol_rate}
\ee
and the copy is thus elongated by one extra nucleotide.  

The reverse reactions on the right-hand side of Fig.~\ref{fig1} are the dissociation of $m_{l+1}{\rm P}$ at the
\be
\mbox{nucleotide dissociation rate:} \qquad k_{-m_{l+1} m_l\atop \ \, n_{l+1} \, n_l} \, ,
\label{nd_rate}
\ee
and the pyrophosphorolysis of the next ultimate unit $m_{l+1}$ of the copy with a pyrophosphate coming from the solution at the 
\be
\mbox{depolymerization rate:} \qquad k^{\rm p}_{-m_{l+1} m_l\atop \ \, n_{l+1} \, n_l}[{\rm P}] \, .
\label{depol_rate}
\ee

The kinetic equations of this scheme are given by Eqs.~(\ref{kin_eq_1})-(\ref{kin_eq_2}) in Appendix~\ref{AppA}.  These equations rule the time evolution of the probabilities
\bea
&&{\cal P}_t\left(m_1 \cdots m_l \qquad\quad\ \atop n_1\, \cdots \, n_l \, n_{l+1}\cdots\right) \qquad\qquad\mbox{and} \nonumber\\
&&{\cal P}_t\left(m_1 \cdots m_l m_{l+1}{\rm P} \qquad\ \ \atop n_1\, \cdots \, n_l \, n_{l+1}\, n_{l+2}\cdots\right) 
\label{probabilities}
\eea
 that the growing copy has respectively the sequences $m_1\cdots m_l$ and $m_1\cdots m_l m_{l+1}{\rm P}$.  These probabilities are proportional to the concentrations of these sequences in a dilute solution.

The sequence $n_1\cdots n_l n_{l+1}\cdots$ of the template $\alpha$ is typically aperiodic and described by a probability distribution $\nu_l(\alpha)=\nu_l(n_1\cdots n_l)$, which is normalized as $\sum_{n_1\cdots n_l}\nu_l(n_1\cdots n_l)=1$. In general, the sequence may have any kind of statistical correlations among the successive monomeric units.  On the one hand, systematic studies have shown that DNA sequences of biological species manifest statistical correlations that cannot be described by low order Markov chains \cite{EFH87,PNN14a,PNN14b}.  On the other hand, arbitrary DNA sequences can be synthesized with modern technologies \cite{CGK12,LOLS13}.  In the following, we assume for simplicity that the template is a Bernoulli chain such that $\nu_l(n_1\cdots n_l)=\prod_{j=1}^l \nu_1(n_j)$ with $\nu_1(n_j)=\frac{1}{4}$ for every $n_j\in\{ {\rm A}, {\rm C}, {\rm G}, {\rm T}\}$.

\subsection{Michaelis-Menten kinetics}
\label{MM_kin}

Experimental observations \cite{LK82,PWJ91,WPJ91,DPJ91,J93,TJ06,JJ01a,JJ01b,LNKC01,LJ06,EG91,KB00,SUKOOBWFWMG08,RBT08,ZBNS09,DJPW10,BBT12} show that the binding and dissociation of $m_{l+1}{\rm P}$ is faster than the incorporation of the nucleotide in the chain:
\be
k_{+m m' \atop \ \, n \, n'}[m{\rm P}], \ k_{-m m'\atop \ \, n \, n'} \gg k^{\rm p}_{+m m' \atop \ \, n \, n'}, \ k^{\rm p}_{-m m'\atop \ \, n \, n'}[{\rm P}] \, .
\label{MM-hyp}
\ee
Accordingly, the molecular chains $m_1\cdots m_l$ and $m_1\cdots m_l m_{l+1}{\rm P}$ are in quasi-equilibrium and the probabilities~(\ref{probabilities}) remain in the following proportionality
\bea
&&\qquad {\cal P}_t\left(m_1 \cdots m_l m_{l+1}{\rm P} \qquad\ \ \atop n_1\, \cdots \, n_l \, n_{l+1}\, n_{l+2}\cdots\right) \nonumber\\
&&\simeq \
 \frac{[m_{l+1}{\rm P}]}{K_{m_{l+1} m_l\atop n_{l+1} \, n_l}} \,
{\cal P}_t\left(m_1 \cdots m_l \qquad\quad\ \atop n_1\, \cdots \, n_l \, n_{l+1}\cdots\right) 
\eea
expressed in terms of the constants
\be
K_{m m'\atop n \, n'}\equiv 
\frac{k_{-m m' \atop \ \, n \, n'}}
{k_{+m m'\atop \ \, n \, n'}}
\label{MM_csts}
\ee
associated with the Michaelis-Menten kinetics \cite{MM13,JG11}.

Consequently, the two kinetic equations~(\ref{kin_eq_1})-(\ref{kin_eq_2}) of Appendix~\ref{AppA} combine together to form the new kinetic equation~(\ref{kin_eq}) for the time evolution of the probability:
\be
P_t(\omega\vert\alpha) = P_t\left(m_1 \cdots m_l \qquad\quad\ \atop n_1\, \cdots \, n_l \, n_{l+1}\cdots\right)
\label{prob}
\ee
defined by the sum~(\ref{prob_sum}) of the probabilities~(\ref{probabilities}).
We thus obtain a simpler kinetics of Michaelis-Menten type with the attachment rate of $m_{l+1}$ given by
\be
W^{\rm p}_{+m_{l+1} m_l\atop \ \, n_{l+1}\, n_l}\equiv 
\frac{k^{\rm p}_{+m_{l+1} m_l\atop \ \, n_{l+1} \, n_l}[m_{l+1}{\rm P}]}
{K_{m_{l+1} m_l\atop n_{l+1} \, n_l}\, Q_{\ \ \ \ \ m_l\atop n_{l+1} \, n_l}}
\label{Wp+}
\ee
and the detachment rate of $m_l$ by
\be
W^{\rm p}_{\quad\, -m_l m_{l-1}\atop n_{l+1}\, n_l \, n_{l-1}}\equiv 
\frac{k^{\rm p}_{-m_l m_{l-1}\atop \ \, n_l \, n_{l-1}}[{\rm P}]}
{Q_{\ \ \ \ \ m_l\atop n_{l+1} \, n_l}} \, ,
\label{Wp-}
\ee
where
\be
Q_{\ \ \ \ \ m_l\atop n_{l+1} \, n_l}\equiv 1 + \sum_{m_{l+1}} \frac{[m_{l+1}{\rm P}]}{K_{m_{l+1} m_l\atop n_{l+1} \, n_l}} \, .
\label{denom}
\ee
The rates~(\ref{Wp+}) and~(\ref{Wp-}) are those of the possible reactive events occurring to the sequence $m_1\cdots m_l$ of the copy on the sequence $n_1\cdots n_l n_{l+1}\cdots$ of the template.  Because of the Michaelis-Menten kinetics, the detachment rate~(\ref{Wp-}) depends not only on the template nucleotides $n_{l-1}$ and $n_l$ forming the base pairs $m_{l-1}$:$n_{l-1}$ and $m_l$:$n_l$, but also on the next template nucleotide $n_{l+1}$.  The stochastic process ruled by the rates~(\ref{Wp+}) and~(\ref{Wp-}) can be numerically simulated with Gillespie's algorithm \cite{G76,G77}, as explained in Appendix~\ref{AppC}.

Experimental data on the rate constants of depolymerization are very rare in the literature.  Data from Ref.~\cite{PWJ91} allows us to infer the depolymerization rate constant in one case, which motivates the assumption that the depolymerization and polymerization rate constants are proportional to each other
\be
k^{\rm p}_{-m_{l+1} m_l\atop \ \, n_{l+1} \, n_l}= \frac{1}{K_{\rm P}} \, k^{\rm p}_{+m_{l+1} m_l\atop \ \, n_{l+1} \, n_l}
\label{K_P-dfn}
\ee
introducing a constant $K_{\rm P}$ associated with pyrophosphorolysis.

Although the concentrations of the four nucleotides may differ in the surrounding solution, they are supposed in the present paper to be all equal to each other:
\be
[{\rm dNTP}] \equiv [{\rm dATP}] = [{\rm dCTP}] = [{\rm dGTP}] = [{\rm dTTP}] \, .
\label{dNTP}
\ee
The analysis of this particular case is simpler, notably because the Michaelis-Menten denominators~(\ref{denom}) reduce to
\be
Q_{\ \ \ \ \ m_l\atop n_{l+1} \, n_l} = 1 + [{\rm dNTP}] \sum_{m_{l+1}} \frac{1}{K_{m_{l+1} m_l\atop n_{l+1} \, n_l}} \, .
\label{denom-dNTP}
\ee

For exonuclease-deficient polymerases, the mean elongation rate, i.e., the mean growth velocity $v$ of the copy, is equal to the production rate $r^{\rm p}$ of pyrophosphate by the reaction~(\ref{pol-react}):
\be
v \equiv \frac{d\langle l\rangle_t}{dt} = r^{\rm p} \, .
\label{velocity}
\ee
The mean growth velocity is vanishing at equilibrium where the polymerase activity stops on average: $v_{\rm eq}=r^{\rm p}_{\rm eq}=0$.

The copolymerization process reaches a regime of steady growth when the mean growth velocity becomes constant in time so that the average length of the copy grows linearly in time \cite{AG08}.  In this regime, the probability~(\ref{prob}) ruled by the kinetic equation~(\ref{kin_eq}) can be written in the form
\bea
P_t(\omega\vert\alpha) &=& P_t\left(m_1 \cdots m_l \qquad\quad\ \ \atop n_1\, \cdots \, n_l \, n_{l+1}\cdots\right) \nonumber\\
&\simeq& p_t(l) \ \underbrace{\mu_l\left(m_1 \cdots m_l \qquad\quad\ \ \atop n_1\, \cdots \, n_l \, n_{l+1}\cdots\right)}_{\mu_l(\omega\vert\alpha)} \nonumber\\
&=& p_t(l) \, \mu_l(\omega\vert\alpha)
\label{prob2}
\eea
in terms of the probability $p_t(l)$ that the copy has the length $l$, and the probability $\mu_l(\omega\vert\alpha)$ that it has the sequence $\omega=m_1 \cdots m_l$ given that its length takes the value~$l$ and the template has the sequence~$\alpha$.  After a long enough time $t\to\infty$, the probability distribution of the length typically behaves as the Gaussian distribution:
\be
p_t(l) \simeq \frac{1}{\sqrt{4\pi {\cal D}t}}\exp\left[-\frac{(l-vt)^2}{4{\cal D}t}\right] \, ,
\label{prob_l}
\ee
where $v>0$ is the mean growth velocity and $\cal D$ a diffusivity coefficient.

Since the formation of incorrect base pairs is in general possible, the copy is not strictly identical to the template.  To characterize this effect caused by molecular fluctuations, we introduce the error probability $\eta$ as the mean number of mismatches per incorporated nucleotide, a mismatch meaning a base pair different from the four Watson-Crick pairs $\{$A:T, C:G, G:C, T:A$\}$.  Out of the sixteen possible pairs $\{m$:$n\}$, four are thus correct and twelve incorrect.

\subsection{Thermodynamics and sequence disorder}

For copolymerization processes, thermodynamics is directly linked to information theory, as previously shown \cite{B79,AG08}.  The basic results used in the following are here summarized and formulated for the present purposes.

If the lapses of time between the reactive events is longer than the relaxation time taken by the DNA molecule to reach thermal equilibrium with the surrounding solution at the temperature $T$, thermodynamic quantities such as the enthalpy $H_l(\omega\vert\alpha)$, the entropy $S_l(\omega\vert\alpha)$, and the free enthalpy $G_l(\omega\vert\alpha)=H_l(\omega\vert\alpha)-TS_l(\omega\vert\alpha)$ can be associated with the copy $\omega$ of length $l$ bounded to the template $\alpha$ and the enzyme.  The average values of these quantities are defined as
\be
\langle X\rangle_t = \sum_{\omega} P_t(\omega\vert\alpha) \, X_l(\omega\vert\alpha) \, .
\ee
The total entropy of the system is given by
\be
S_t = \sum_{\omega} P_t(\omega\vert\alpha) \, S_l(\omega\vert\alpha) - R \sum_{\omega} P_t(\omega\vert\alpha) \, \ln P_t(\omega\vert\alpha) \, ,
\label{entr}
\ee
where the first contribution comes from disorder in the internal degrees of freedom and the second from disorder among the population of different sequences $\omega$ that are possible at a given time $t$ \cite{P55,S76,N79,LVN84,JQQ04,G04}.  An expression similar to Eq.~(\ref{entr}) is obtained in terms of the concentrations of the different sequences in a dilute solution where the concentrations are proportional to the probabilities.

Now, the different thermodynamic quantities  change in time because the probabilities~(\ref{prob}) have a time evolution ruled by the kinetic equation~(\ref{kin_eq}).  In particular, the balance of entropy can be established as for general reactive processes and the entropy production can be obtained, which is given by Eq.~(\ref{entr-A}) in Appendix~\ref{AppA}.  In the regime of steady growth, the entropy production -- which is always non-negative by the second law of thermodynamics -- can be written as \cite{AG08}
\be
\Sigma \equiv \frac{1}{R}\frac{d_{\rm i}S}{dt}= v \, A \geq 0 \, ,
\label{entr-prod}
\ee
in terms of the mean growth velocity in nucleotides per second and the entropy production per nucleotide, also called affinity:
\be
A = \epsilon + D(\omega\vert\alpha) \, .
\label{A}
\ee
This latter has two contributions.  The first one is the mean free-energy driving force per nucleotide
\be
\epsilon = \lim_{L\to\infty} \frac{1}{L} \sum_{l=1}^L \epsilon_l
\label{eps}
\ee
with
\be
\epsilon_l \equiv \ln \left(W^{\rm p}_{+m_l m_{l-1}\atop \ \, n_l\, n_{l-1}}/
W^{\rm p}_{\quad\, -m_l m_{l-1}\atop n_{l+1}\, n_l \, n_{l-1}}\right) \, .
\label{eps_l}
\ee
This driving force can be expressed as $\epsilon=-g/(RT)$ in terms of the free enthalpy per nucleotide $g$ incorporated in the copy, which is negative if the attachment is energetically favorable because the free energy landscape is thus going down in the direction of growth.  The second contribution in Eq.~(\ref{A}) is the conditional Shannon disorder per nucleotide of the copy $\omega$ with respect to the template $\alpha$:
\be
D(\omega\vert\alpha) = \lim_{l\to\infty} -\frac{1}{l} \sum_{\alpha,\omega} \nu_l(\alpha)\, \mu_l(\omega\vert\alpha) \, \ln \mu_l(\omega\vert\alpha) \geq 0 \, .
\label{Dc}
\ee
If fidelity is high in the replication process, the vast majority of copies $\omega$ are identical to the template $\alpha$ and the errors are thus very rare.  In this case, the errors may be assumed to be statistically independent of each other.  If moreover the substitutions are equiprobable, which is favored by equal nucleotide concentrations~(\ref{dNTP}), the conditional disorder can be estimated as
\be
D(\omega\vert\alpha) \simeq \eta \, \ln \frac{3{\rm e}}{\eta} 
\label{D-estim}
\ee
in terms of the error probability $\eta \ll 1$.  In view of Eqs.~(\ref{entr-prod}) and~(\ref{A}), the error probability is thus related to the  thermodynamic entropy production \cite{AG08}.

Using information theory \cite{CT06}, the conditional disorder~(\ref{Dc}) can be expressed as
\be
D(\omega\vert\alpha) = D(\omega)- I(\omega,\alpha)
\label{Dc=D-I}
\ee
in terms of the overall Shannon disorder $D(\omega)$ of the copy and
the mutual information $I(\omega,\alpha)$ between the copy and the template \cite{AG08}.  The mutual information characterizes the fidelity of the copying process. The larger the mutual information, the higher the fidelity of DNA replication.

The overall disorder per nucleotide is defined by
\be
D(\omega) = \lim_{l\to\infty} -\frac{1}{l} \sum_{\omega} \mu_l(\omega) \, \ln \mu_l(\omega) \geq 0 \, ,
\label{D}
\ee
where $\mu_l(\omega) = \sum_{\alpha} \nu_l(\alpha)\, \mu_l(\omega\vert\alpha)$ is the probability distribution of the copy for any template sequence.  This quantity, which is often called information Shannon entropy, is studied to characterize the complexity of DNA symbolic sequences in various biological organisms \cite{EFH87,PNN14a,PNN14b}.  The overall disorder per nucleotide is observed to vary from the value $D(\alpha)\simeq 1.339$ assuming the sequence $\alpha$ is a first-order Markov chain, down to $D(\alpha)\simeq 1.273$ if $\alpha$ is a $8^{\rm th}$-order Markov chain, which suggests the existence of long-range correlations besides the fact that the four nucleotides occur with unequal probabilities in typical DNA sequences \cite{PNN14a,PNN14b}.  In the present paper, the template is supposed to be a Bernoulli chain with equal probabilities $\nu_1(n)=\frac{1}{4}$ for the different nucleotides $n\in\{{\rm A},{\rm C},{\rm G},{\rm T}\}$, so that the overall disorders of the template and the copy reach their maximal values $D(\alpha)=D(\omega)=\ln 4\simeq 1.386$.  

It should be pointed out that the sequence carries information to the extent that it is replicated, transcripted, or translated into proteins in living organisms.  {\it A priori}, the sequences of the template $\alpha$ and the copy $\omega$ only appear disordered.  It is the fidelity of the copying process that allows these sequences to acquire meaning. In this regard, it is the mutual information~$I(\omega,\alpha)$ that is specific to the replication of genetic information.  If the coupling was loose between the copy and the template, the copolymerization would be free from the template and the mutual information would vanish.  For a tight coupling, the error probability is expected to take a small value $\eta\ll 1$, as well as the conditional disorder~(\ref{D-estim}).  In this case, the mutual information between the copy and the template can be estimated as
\be
I(\omega,\alpha) \simeq \ln 4 - \eta \, \ln \frac{3{\rm e}}{\eta} \, ,
\label{I-estim}
\ee
which is very close to its maximal value ${\rm Max}\{I(\omega,\alpha)\}=\ln 4$.

At equilibrium, the thermodynamic entropy production~(\ref{entr-prod}) is vanishing with the velocity~(\ref{velocity}) and the affinity~(\ref{A}).  Accordingly, the equilibrium value of the free-energy driving force~(\ref{eps})-(\ref{eps_l}) is fully determined by the conditional Shannon disorder and the error probability: 
 \be
 \epsilon_{\rm eq}=-D(\omega\vert\alpha)_{\rm eq}\simeq -\eta_{\rm eq}\ln(3{\rm e}/\eta_{\rm eq}) \, .
 \label{eps-eq}
 \ee

\section{Bernoulli-chain model}
\label{B-chain}

\subsection{Kinetics and error probability}

The stochastic process introduced in Subsection~\ref{MM_kin} can be compared with simplified models, which are analytically solvable.  The simplest one is based on the two following assumptions that the rates do not depend on the previously incorporated nucleotide and, moreover, that the rates only depend on whether the pairing is correct or incorrect.  Although the first assumption is not supported by experimental observations~\cite{J93}, it is often considered because of its great simplicity.  The second assumption captures the observation that the polymerization rate constants $k^{\rm p}_{+m\atop \ \, n}$ and the Michaelis-Menten constants $K_{m\atop n}$ defined by Eqs.~(\ref{MM_csts}) take similar values within the set of correct (respectively incorrect) pairings \cite{LK82,PWJ91,WPJ91,DPJ91,J93,TJ06,JJ01a,JJ01b,LNKC01,LJ06,EG91,KB00,SUKOOBWFWMG08,RBT08,ZBNS09,DJPW10,BBT12}.  According to these assumptions, the model only needs the four rate constants $k^{\rm p}_{\pm{\rm c}}$ and $k^{\rm p}_{\pm{\rm i}}$ for polymerization and depolymerization, together with the two Michaelis-Menten constants $K_{\rm c}$ and $K_{\rm i}$ for correct and incorrect pairings.  The simplification that consists in reducing the description to correct and incorrect pairings is also often used to study DNA replication \cite{B79,CQ09,SC12,SP13,S14,MHL12,MHL14,RP15}.

In spite of their essential role in establishing the thermodynamics of DNA polymerase activity, there are very few experimental data published in the literature on the depolymerization rate constants $k^{\rm p}_{-m\atop \ \, n}$.  The experimental measurements reported in Ref.~\cite{PWJ91} gives us the ratio between the polymerization to the depolymerization rate constants:
\be
K_{\rm P}\equiv\frac{k^{\rm p}_{+{\rm c}}}{k^{\rm p}_{-{\rm c}}}=\frac{k^{\rm p}_{+{\rm i}}}{k^{\rm p}_{-{\rm i}}} \, .
\label{K_P-B}
\ee
Although this knowledge is limited, it allows us to determine the depolymerization rate constants $k^{\rm p}_{-m\atop \ \, n}$ of pyrophosphorolysis in terms of the well-known polymerization rate constants $k^{\rm p}_{+m\atop \ \, n}$, which is essential for the chemical equilibrium thermodynamics of DNA polymerase activity.

Furthermore, it is supposed that the concentrations of the four nucleotides are equal in the surrounding solution, as expressed by Eq.~(\ref{dNTP}).

Under these assumptions, the model is defined by the attachment and detachment rates:
\bea
&&W^{\rm p}_{+{\rm c}} = \frac{k^{\rm p}_{+{\rm c}} \, [{\rm dNTP}]}{K_{\rm c} Q} \, , \quad
W^{\rm p}_{+{\rm i}} = \frac{k^{\rm p}_{+{\rm i}} \, [{\rm dNTP}]}{K_{\rm i} Q} \, , \\
&&W^{\rm p}_{-{\rm c}} = \frac{k^{\rm p}_{+{\rm c}} \, [{\rm P}]}{K_{\rm P} Q} \, , \qquad\quad
W^{\rm p}_{-{\rm i}} = \frac{k^{\rm p}_{+{\rm i}} \, [{\rm P}]}{K_{\rm P} Q}\, ,
\eea
with the Michaelis-Menten denominator:
\be
Q = 1 + \left(\frac{1}{K_{\rm c}}+\frac{3}{K_{\rm i}}\right) [{\rm dNTP}] \, .
\ee

For the so-defined model, the process is similar to the simplest free copolymerization, which is exactly solvable \cite{AG09}.  The growing copy is a Bernoulli chain, whereupon the probability of a sequence $\omega$ factorizes as
\bea
\mu_l(\omega\vert\alpha) &=& \mu_l\left(m_1 m_2\cdots m_l \qquad\quad\ \ \atop n_1\, n_2 \, \cdots \, n_l \, n_{l+1}\cdots\right) \nonumber\\
&=& \mu(p_1)\, \mu(p_2) \cdots \mu(p_l)
\eea
in terms of the probabilities
\be
\mu(p_j)\equiv\mu_1\left(m_j\atop n_j\right) \qquad \mbox{with} \quad p_j={\rm c} \ \ \mbox{or}\ \ {\rm i}
\ee
that the base pair $p_j$ is correct or incorrect.  These probabilities are given by
\bea
\mu({\rm c}) &=& \frac{W^{\rm p}_{+{\rm c}}}{W^{\rm p}_{-{\rm c}}+v}  \, , \label{B_mu_c}\\
\mu({\rm i}) &=& \frac{W^{\rm p}_{+{\rm i}}}{W^{\rm p}_{-{\rm i}}+v}  \, , \label{B_mu_i}
\eea
where $v$ is the mean growth velocity. Because of the normalization condition
\be
\mu({\rm c}) + 3 \, \mu({\rm i}) = 1 \, ,
\ee
the error probability is here defined by
\be
\eta \equiv 1-\mu({\rm c}) = 3 \, \mu({\rm i}) \, .
\ee
Consequently, the mean growth velocity can be expressed as
\be
v = \frac{W^{\rm p}_{+{\rm c}}}{1-\eta} - W^{\rm p}_{-{\rm c}} = 3\, \frac{W^{\rm p}_{+{\rm i}}}{\eta} - W^{\rm p}_{-{\rm i}}
\label{B-velocity}
\ee
in terms of the error probability $\eta$, providing a closed equation for this latter, which is thus given by the positive root of a quadratic polynomial.

\subsection{Thermodynamics and sequence disorder}

The thermodynamic entropy production is given by Eqs.~(\ref{entr-prod})-(\ref{A}) with the free-energy driving force per nucleotide
\be
\epsilon = (1-\eta) \, \ln \frac{W^{\rm p}_{+{\rm c}}}{W^{\rm p}_{-{\rm c}}}+ \eta \, \ln \frac{W^{\rm p}_{+{\rm i}}}{W^{\rm p}_{-{\rm i}}} \, ,
\label{T.e}
\ee
and the conditional Shannon disorder per nucleotide
\bea
D(\omega\vert\alpha) &=& - \mu({\rm c}) \, \ln\mu({\rm c}) + 3 \, \mu({\rm i}) \, \ln\mu({\rm i}) 
\nonumber\\
&=& -(1-\eta)\, \ln(1-\eta) -\eta \, \ln\frac{\eta}{3} \, ,
\label{D-B}
\eea
as it should for a Bernoulli chain of probabilities $\{1-\eta,\frac{\eta}{3},\frac{\eta}{3},\frac{\eta}{3}\}$.

If the error probability is very small $\eta \ll 1$, the conditional Shannon disorder can be evaluated by Eq.~(\ref{D-estim}), and the mutual information between the copy and the template by Eq.~(\ref{I-estim}) if the template is also a Bernoulli chain.  

An important issue is to determine how the overall sequence disorder evolves during replication.  The Bernoulli-chain model allows us to obtain the overall disorder~(\ref{D}) of the copy $\omega$ in terms of the overall disorder $D(\alpha)$ of the template.  If this latter is a Bernoulli chain of probabilities $\nu(n)\equiv\nu_1(n)=\frac{1}{4}+\delta\nu(n)$ with $\sum_n \delta\nu(n)=0$ and $\vert\delta\nu(n)\vert\ll\frac{1}{4}$, its overall disorder per nucleotide is estimated as $D(\alpha)=-\sum_n\nu(n)\ln\nu(n)\simeq\ln 4 -2\Delta_2$ with $\Delta_2=\sum_n\delta\nu(n)^2$.  After replication, the copy is itself a Bernoulli chain of probabilities given by
\be
\mu(m)=\sum_n \nu(n)\, \mu\left(m\atop n\right) = \left( 1 -\frac{4\, \eta}{3}\right) \nu(\tilde m) + \frac{\eta}{3}
\ee
in terms of the error probability $\eta$, where $\tilde m$ denotes the nucleotide complementary to $m$.  Consequently, the overall disorder per nucleotide of the copy is given by
\be
D(\omega) - \ln 4 \simeq \left( 1 -\frac{4\, \eta}{3}\right)^2 \left[D(\alpha)-\ln 4\right] \, ,
\label{Dw-Da}
\ee
which implies an increase of the overall disorder towards its maximal value ${\rm Max}\{D(\omega)\}=\ln 4$.  After $N$ successive replications, the overall disorder $D_N$ would thus increase as
\be
D_N \simeq \ln 4 - (\ln 4 - D_0) \, \exp\left(-\frac{8\, \eta}{3} \, N\right)
\label{Dw-Da-N}
\ee  
from its initial value $D_0$, if $\eta \ll 1$.  If the nucleotides have equal probabilities so that $\Delta_2=0$, we notice that the overall disorder remains constant from generation to generation: $D(\omega)=D(\alpha)=\ln 4$.

\vskip 0.3 cm

Now, we shall directly obtain the error probability in terms of the rate constants at equilibrium, as well as in the full speed regime.

\subsection{Equilibrium}

If the polymerase activity is at thermodynamic equilibrium, the growth velocity is vanishing $v_{\rm eq}=0$, so that Eqs.~(\ref{B-velocity}) give us two equations for the error probability.  They determine the critical value of dNTP concentration where the equilibrium happens, as well as the error probability at equilibrium:
\bea
&&[{\rm dNTP}]_{\rm eq,B} = \frac{[{\rm P}]}{K_{\rm P}}\left(\frac{1}{K_{\rm c}}+\frac{3}{K_{\rm i}}\right)^{-1} \, , \label{dNTP-eq-B}\\
&& \eta_{\rm eq,B} = \left( 1+\frac{K_{\rm i}}{3\, K_{\rm c}}\right)^{-1} \, , \label{eta-eq-B0}
\eea
 in the Bernoulli-chain model.
 
Since the Michaelis-Menten constant is typically larger for incorrect than correct pairing $K_{\rm i}\gg K_{\rm c}$, the equilibrium error probability is well approximated by
 \be
  \eta_{\rm eq,B} \simeq 3\, \frac{K_{\rm c}}{K_{\rm i}} \ll 1 \, .
 \label{eta-eq-B}
 \ee

The equilibrium values of the conditional Shannon disorder and the free-energy driving force are determined in terms of the equilibrium error probability by Eq.~(\ref{eps-eq}).
  
We notice that the interesting approximation
 \be
 v\simeq k^{\rm p}_{+{\rm c}}\, \frac{\left(\frac{1}{K_{\rm c}}+\frac{3}{K_{\rm i}}\right) [{\rm dNTP}]-\frac{1}{K_{\rm P}}\, [{\rm P}]}{1 + \left(\frac{1}{K_{\rm c}}+\frac{3}{K_{\rm i}}\right) [{\rm dNTP}]}
 \label{MM-velocity}
 \ee
can be obtained for the growth velocity after substituting the equilibrium error probability~(\ref{eta-eq-B0}) into the first of expressions~(\ref{B-velocity}).  This approximation explicitly shows that the polymerase activity is ruled by a Michaelis-Menten kinetics and the growth velocity vanishes at the critical concentration~(\ref{dNTP-eq-B}).

\subsection{Full speed regime}

The full speed regime of the enzyme is reached if the substrate concentrations are larger than the Michaelis-Menten constant
\be
[{\rm dNTP}] \gg \left(\frac{1}{K_{\rm c}}+\frac{3}{K_{\rm i}}\right)^{-1} \, .
\ee
In this regime, the detachment rates become negligible ($W^{\rm p}_{-{\rm c}}, W^{\rm p}_{-{\rm i}} \ll W^{\rm p}_{+{\rm c}}, W^{\rm p}_{+{\rm i}}$) and Eqs.~(\ref{B-velocity}) give us the mean growth velocity and the error probability as
\bea
&& v_{\infty,{\rm B}} = \frac{k^{\rm p}_{+{\rm c}} \, K_{\rm i} + 3 \, k^{\rm p}_{+{\rm i}} \, K_{\rm c}}{K_{\rm i} + 3 \, K_{\rm c}} \, ,\label{v-infty-B0}
\\
&& \eta_{\infty,{\rm B}} = \left( 1+\frac{k^{\rm p}_{+{\rm c}} \, K_{\rm i}}{3 \, k^{\rm p}_{+{\rm i}} \, K_{\rm c}}\right)^{-1} \, . \label{eta-infty-B0}
\eea

Since the polymerization rate constant is typically larger for correct than incorrect pairing $k^{\rm p}_{+{\rm c}}\gg k^{\rm p}_{+{\rm i}}$ while the Michaelis-Menten dissociation constant is smaller $K_{\rm c}\ll K_{\rm i}$, the growth velocity and the error probability can be approximated at full speed by
\bea
&& v_{\infty,{\rm B}} \simeq k^{\rm p}_{+{\rm c}} \, ,\label{v-infty-B}
\\
&&\eta_{\infty,{\rm B}} \simeq 3\, \frac{k^{\rm p}_{+{\rm i}}\, K_{\rm c}}{k^{\rm p}_{+{\rm c}}\, K_{\rm i}} \, .
\label{eta-infty-B}
\eea

If the velocity and the error probability reach a plateau as the dNTP concentration increases in the full speed regime, the thermodynamic entropy production~(\ref{entr-prod}) instead roughly increases as
\be
\frac{1}{R}\frac{d_{\rm i}S}{dt} \simeq v_{\infty,{\rm B}} \, \ln\frac{K_{\rm P}[{\rm dNTP}]}{K_{\rm c}[{\rm P}]}
\label{entr-prod-infty-B}
\ee
with the concentration [dNTP], if the error probability is so small that its effects become negligible.  Under the same conditions, the affinity and free-energy driving force per nucleotide increase as
\be
A \simeq \epsilon \simeq \ln\frac{K_{\rm P}[{\rm dNTP}]}{K_{\rm c}[{\rm P}]}
\label{A-infty-B}
\ee
with the concentration [dNTP].

\section{Markov-chain model}
\label{M-chain}

\subsection{Kinetics and error probability}

An essential aspect of DNA polymerases is that their rates depend not only on the nucleotide that is attached or detached, but also on the previously incorporated nucleotide, because the enzyme is sensitive to mismatches~\cite{J93}.  Accordingly, the assumptions of the Bernoulli-chain model are too restrictive and we need to extend the model.  As before, we suppose that the kinetic constants only depend on whether the pairing is correct or incorrect without further distinction.  Therefore, the attachment rates~(\ref{Wp+}) are here defined by
\bea
W^{\rm p}_{+{\rm c}\vert{\rm c}} = \frac{k^{\rm p}_{+{\rm c}\vert{\rm c}}[{\rm dNTP}]}{K_{{\rm c}\vert{\rm c}}Q_{\rm c}} \, , && W^{\rm p}_{+{\rm i}\vert{\rm c}} = \frac{k^{\rm p}_{+{\rm i}\vert{\rm c}}[{\rm dNTP}]}{K_{{\rm i}\vert{\rm c}}Q_{\rm c}} \, , \label{M_W_c}\\
W^{\rm p}_{+{\rm c}\vert{\rm i}} = \frac{k^{\rm p}_{+{\rm c}\vert{\rm i}}[{\rm dNTP}]}{K_{{\rm c}\vert{\rm i}}Q_{\rm i}} \, , && W^{\rm p}_{+{\rm i}\vert{\rm i}} = \frac{k^{\rm p}_{+{\rm i}\vert{\rm i}}[{\rm dNTP}]}{K_{{\rm i}\vert{\rm i}}Q_{\rm i}} \, , \label{M_W_i}
\eea
with the denominators
\bea
&& Q_{\rm c} = 1 + \left(\frac{1}{K_{{\rm c}\vert{\rm c}}}+\frac{3}{K_{{\rm i}\vert{\rm c}}}\right) [{\rm dNTP}] \, , 
\label{Qc}\\
&& Q_{\rm i} = 1 + \left(\frac{1}{K_{{\rm c}\vert{\rm i}}}+\frac{3}{K_{{\rm i}\vert{\rm i}}}\right) [{\rm dNTP}] \, ,
\label{Qi}
\eea
obtained form Eq.~(\ref{denom-dNTP}).  We notice that these denominators no longer depend on the template nucleotide $n_{l+1}$ because the Michaelis-Menten dissociation constants are supposed to differ only between correct and incorrect pairings and the nucleotide concentrations are taken equal to each other by Eq.~(\ref{dNTP}).

As before, the detachment rate constants are determined from the knowledge of the constant associated with pyrophosphorolysis:
\be
K_{\rm P}\equiv\frac{k^{\rm p}_{+{\rm c}\vert{\rm c}}}{k^{\rm p}_{-{\rm c}\vert{\rm c}}}=\frac{k^{\rm p}_{+{\rm i}\vert{\rm c}}}{k^{\rm p}_{-{\rm i}\vert{\rm c}}}=\frac{k^{\rm p}_{+{\rm c}\vert{\rm i}}}{k^{\rm p}_{-{\rm c}\vert{\rm i}}}=\frac{k^{\rm p}_{+{\rm i}\vert{\rm i}}}{k^{\rm p}_{-{\rm i}\vert{\rm i}}} \, .
\label{K_P-M}
\ee
The detachment rates~(\ref{Wp-}) are thus given by
\bea
&&W^{\rm p}_{-{\rm c}\vert{\rm c}} = \frac{k^{\rm p}_{+{\rm c}\vert{\rm c}}[{\rm P}]}{K_{\rm P}Q_{\rm c}} \, , \qquad W^{\rm p}_{-{\rm i}\vert{\rm c}} = \frac{k^{\rm p}_{+{\rm i}\vert{\rm c}}[{\rm P}]}{K_{\rm P}Q_{\rm i}} \, , \\
&&W^{\rm p}_{-{\rm c}\vert{\rm i}} = \frac{k^{\rm p}_{+{\rm c}\vert{\rm i}}[{\rm P}]}{K_{\rm P}Q_{\rm c}} \, , \qquad\ W^{\rm p}_{-{\rm i}\vert{\rm i}} = \frac{k^{\rm p}_{+{\rm i}\vert{\rm i}}[{\rm P}]}{K_{\rm P}Q_{\rm i}} \, , 
\eea
with the denominators~(\ref{Qc}) and~(\ref{Qi}).

For this model, the process is analogous to another free copolymerization process, which is also exactly solvable as recently shown \cite{GA14}.  The growing copy is now a Markov chain, in which case the sequence probability factorizes as
\bea
\mu_l(\omega\vert\alpha) &=& \mu_l\left(m_1 m_2\cdots m_l \qquad\quad\ \ \atop n_1\, n_2 \, \cdots \, n_l \, n_{l+1}\cdots\right) \nonumber\\
&=& \mu(p_1\vert p_2) \cdots \mu(p_{l-1}\vert p_l) \, \mu(p_l)
\label{Markov_chain}
\eea
with $p_j\in\{{\rm c},{\rm i},{\rm i},{\rm i}\}$.  Here, we have introduced the conditional probabilities that a base pair is $p$ given that the next one is $p'$:
\be
\mu(p\vert p')\equiv\mu\left({m\atop n}\bigg\vert{m'\atop n' }\right)
\label{M_cond_prob}
\ee
and the tip probabilities, i.e., the probabilities that the ultimate base pair is $p=$~c or~i:
\be
\mu(p)\equiv\mu\left(m\atop n\right) \, .
\label{M_tip_prob}
\ee

The method of Ref.~\cite{GA14} can be adapted as shown in Appendix~\ref{AppB} in order to calculate these probabilities.  In general, the tip probabilities $\mu(p)$ differ from the bulk probabilities $\bar\mu(p)$ given by the stationary probabilities of the Markov chain:
\be
\sum_{p'} \mu(p\vert p') \, \bar\mu(p') = \bar\mu(p) \, .
\label{M_bulk_prob}
\ee
The tip and bulk probabilities satisfy the normalization conditions:
\bea
\mu({\rm c}) +3 \, \mu({\rm i}) = 1 \, , \label{norm_tip_prob}\\
\bar\mu({\rm c}) +3 \, \bar\mu({\rm i}) = 1 \, . \label{norm_bulk_prob}
\eea

For the Markov chain, the error probability is defined in terms of the bulk probabilities as
\be
\eta \equiv 1-\bar\mu({\rm c}) = 3 \, \bar\mu({\rm i}) \, .
\label{M_error}
\ee

Partial velocities are introduced as
\be
v_p \equiv v \, \frac{\bar\mu(p)}{\mu(p)} \qquad\mbox{for} \quad p\in\{{\rm c},{\rm i},{\rm i},{\rm i}\} \, .
\label{M_part_velocities}
\ee
in terms of the mean growth velocity $v$, the bulk, and the tip probabilities.  The partial velocities can be calculated directly from the knowledge of the transition rates~\cite{GA14}.
The mean growth velocity can then be obtained by averaging the partial velocities over the tip probability distribution:
\be
v = v_{\rm c} \, \mu({\rm c}) + 3 \, v_{\rm i} \, \mu({\rm i}) \, .
\label{M_velocity}
\ee
Further details are given in Appendix~\ref{AppB}.

\subsection{Thermodynamics and sequence disorder}

For the Markov-chain model, the thermodynamic entropy production is also given by Eqs.~(\ref{entr-prod})-(\ref{A}), but with the free-energy driving force per nucleotide
\bea
\epsilon &=& \bar\mu({\rm c}) \, \mu({\rm c}\vert{\rm c}) \, \ln \frac{W^{\rm p}_{+{\rm c}\vert{\rm c}}}{W^{\rm p}_{-{\rm c}\vert{\rm c}}}+ 3\, \bar\mu({\rm c}) \, \mu({\rm i}\vert{\rm c}) \, \ln \frac{W^{\rm p}_{+{\rm c}\vert{\rm i}}}{W^{\rm p}_{-{\rm c}\vert{\rm i}}} \nonumber\\
&+& 3 \, \bar\mu({\rm i}) \, \mu({\rm c}\vert{\rm i}) \, \ln \frac{W^{\rm p}_{+{\rm i}\vert{\rm c}}}{W^{\rm p}_{-{\rm i}\vert{\rm c}}}+ 9\, \bar\mu({\rm i}) \, \mu({\rm i}\vert{\rm i}) \, \ln \frac{W^{\rm p}_{+{\rm i}\vert{\rm i}}}{W^{\rm p}_{-{\rm i}\vert{\rm i}}} \, , \quad
\label{eps-M}
\eea
and the conditional Shannon disorder per nucleotide
\bea
D(\omega\vert\alpha) &=& \, -\,\bar\mu({\rm c}) \, \mu({\rm c}\vert{\rm c}) \, \ln \mu({\rm c}\vert{\rm c}) \nonumber\\
&& - 3\, \bar\mu({\rm c}) \, \mu({\rm i}\vert{\rm c}) \, \ln \mu({\rm i}\vert{\rm c}) 
\nonumber\\
&& - \, 3 \, \bar\mu({\rm i}) \, \mu({\rm c}\vert{\rm i}) \, \ln\mu({\rm c}\vert{\rm i}) 
\nonumber\\
&& - 9\, \bar\mu({\rm i}) \, \mu({\rm i}\vert{\rm i}) \, \ln\mu({\rm i}\vert{\rm i}) \, , 
\label{D-M}
\eea
as it should for a Markov chain \cite{GA14}.

For a very small error probability $\eta \ll 1$, the conditional Shannon disorder~(\ref{D-M}) can again be evaluated by Eq.~(\ref{D-estim}), and the mutual information by Eq.~(\ref{I-estim}) if the template is a Bernoulli chain.  

\subsection{Equilibrium}

At equilibrium, the mean and partial velocities are vanishing, $v=v_{\rm c}=v_{\rm i}=0$, together with the entropy production~(\ref{entr-prod}) and the affinity~(\ref{A}). Typically, the Michaelis-Menten constants of DNA polymerases are ordered as $K_{{\rm c}\vert{\rm c}}\ll K_{{\rm i}\vert{\rm c}},K_{{\rm i}\vert{\rm i}}$.  As shown in Appendix~\ref{AppB}, the error probability can be evaluated in this case as
\be
\eta_{\rm eq,M} \simeq 3 \, \frac{K_{{\rm c}\vert{\rm c}}^2}{K_{{\rm c}\vert{\rm i}}K_{{\rm i}\vert{\rm c}}} \ll 1 \, .
\label{eta-eq-M}
\ee
We notice that the error probability~(\ref{eta-eq-B}) is recovered for the Bernoulli-chain model where $K_{{\rm c}\vert{\rm c}}=K_{{\rm c}\vert{\rm i}}=K_{\rm c}$ and $K_{{\rm i}\vert{\rm c}}=K_{{\rm i}\vert{\rm i}}=K_{\rm i}$.  However, the equilibrium error probability~(\ref{eta-eq-M}) of the Markov-chain model can take significantly lower values than in the Bernoulli-chain model if moreover $K_{{\rm c}\vert{\rm c}}\ll K_{{\rm c}\vert{\rm i}}$.

For the polymerase activity, the mean growth velocity is vanishing at the thermodynamic equilibrium concentration:
\be
[{\rm dNTP}]_{\rm eq,M}= \frac{[{\rm P}]}{K_{\rm P}} \, K_{{\rm c}\vert{\rm c}} \, (1+\delta) \qquad\mbox{with}\quad \delta\simeq - \eta_{\rm eq,M} \, ,
\label{dNTP-eq-M}
\ee
which is also shown in Appendix~\ref{AppB}.

Again, the equilibrium free-energy driving force is related to the conditional disorder and the error probability by Eq.~(\ref{eps-eq}).

\subsection{Full speed regime}

The full speed regime is reached if the nucleotide concentrations satisfy the conditions
\be
[{\rm dNTP}] \gg \left(\frac{1}{K_{{\rm c}\vert p}}+\frac{3}{K_{{\rm i}\vert p}}\right)^{-1}
\ee
for $p=$~c and~i.
In this regime, the detachment rates become negligible with respect to the attachment rates. Moreover, the attachment rate of a correct base pair after the incorporation of a correct base pair is typically larger than the other ones.  In such circumstances, the mean growth velocity can be evaluated as
\be
v_{\infty,{\rm M}} \simeq \frac{k^{\rm p}_{+{\rm c}\vert{\rm c}}}
{\displaystyle 1+3\,\frac{K_{{\rm c}\vert{\rm c}}}{K_{{\rm i}\vert{\rm c}}}}
\label{v-infty-M0}
\ee
and the error probability as
\be
\eta_{\infty,{\rm M}} \simeq 3 \, \frac{k^{\rm p}_{+{\rm i}\vert{\rm c}} K_{{\rm c}\vert{\rm c}}}{k^{\rm p}_{+{\rm c}\vert{\rm c}}K_{{\rm i}\vert{\rm c}}}\left( 1+ 3\, \frac{k^{\rm p}_{+{\rm i}\vert{\rm i}}  K_{{\rm c}\vert{\rm i}}}{k^{\rm p}_{+{\rm c}\vert{\rm i}} K_{{\rm i}\vert{\rm i}}}\right) \, ,
\label{eta-infty-M0}
\ee
as explained in Appendix~\ref{AppB}.

If the polymerization rate constants are larger for correct than incorrect incorporation and the Michaelis-Menten dissociation constants smaller, the growth velocity and the error probability can be approximated by
\bea
&&v_{\infty,{\rm M}} \simeq k^{\rm p}_{+{\rm c}\vert{\rm c}} \, , \label{v-infty-M} \\
&&\eta_{\infty,{\rm M}} \simeq 3 \, \frac{k^{\rm p}_{+{\rm i}\vert{\rm c}} K_{{\rm c}\vert{\rm c}}}{k^{\rm p}_{+{\rm c}\vert{\rm c}}K_{{\rm i}\vert{\rm c}}} \, ,
\label{eta-infty-M}
\eea
which are similar to the expressions~(\ref{v-infty-B}) and~(\ref{eta-infty-B}) for the Bernoulli-chain model.

In the full speed regime, the affinity and the free-energy driving force per nucleotide are nearly equal if the error probability~(\ref{eta-infty-M}) is very small and they increases as
\be
A \simeq \epsilon \simeq \ln\frac{K_{\rm P}[{\rm dNTP}]}{K_{{\rm c}\vert{\rm c}}[{\rm P}]}
\label{A-infty-M}
\ee
with the concentration [dNTP]. Therefore, the thermodynamic entropy production~(\ref{entr-prod}) also increases as the logarithm of the dNTP concentration
\be
\frac{1}{R}\frac{d_{\rm i}S}{dt} \simeq v_{\infty,{\rm M}} \, \ln\frac{K_{\rm P}[{\rm dNTP}]}{K_{{\rm c}\vert{\rm c}}[{\rm P}]} \, ,
\label{entr-prod-infty-M}
\ee
since the mean growth velocity saturates at the plateau value~(\ref{v-infty-M0}).  In the full speed regime, the behavior is similar as in the Bernoulli-chain model.

The results (\ref{A-infty-M})-(\ref{entr-prod-infty-M}) show the importance of knowing the constant $K_{\rm P}$, which characterizes pyrophosphorolysis at equilibrium, in order to determine the (nonequilibrium) thermodynamics of polymerase activity.

\section{T7 DNA polymerase}
\label{T7-Pol}

\subsection{Phenomenology}

The DNA polymerase of the virus phage T7 is a complex of two proteins: the phage protein (80 kDa) and the host {\it E. coli} accessory protein (12 kDa) \cite{PWJ91}.  The phage protein contains both the polymerase and exonuclease activities in the wild type, but the exonuclease activity is suppressed in the exo$^-$ mutant used in the detailed kinetic studies reported in Refs.~\cite{PWJ91,WPJ91,J93}.  The first paper~\cite{PWJ91} is focused on the kinetics of correct nucleotide incorporation, while the second paper~\cite{WPJ91} on incorrect nucleotide incorporation.  The experimentally measured values of these papers have been compiled in Ref.~\cite{J93} and are given in Table~\ref{tab.T7-simul}.  A complete set of rate constants is not available for all possible base pairs, but only for correct and incorrect pairings.  

The key observation is that the constants significantly depend on whether the previously incorporated nucleotide is correct or incorrect \cite{WPJ91,J93}, so that the Markov-chain model applies, but the Bernoulli one does not.  Here, we shall compare the properties of the two models in order to better understand their consequences.  The parameters of a Bernoulli-chain model inferred from the experimental data are given in Table~\ref{tab.T7-B}.  

In order to obtain the thermodynamic quantities, we need data about the transitions that are running backward with respect to the elongation of DNA, in particular, about the pyrophosphorolysis of the DNA growing end.  Experimental data are sparse on the rate constants of these reactions, but Ref.~\cite{PWJ91} provides us with the equilibrium constant of the overall reaction for the correct nucleotide incorporation, from which we infer the value of the constant introduced in Eq.~(\ref{K_P-dfn}): $K_{\rm P}=200$~mM.  In the following, we use the value $[{\rm P}]=10^{-4}$~M for the pyrophosphate concentration, which corresponds to physiological conditions \cite{H01}.

The dissociation rate of the enzyme-DNA complex in Eq.~(\ref{E+DNA}) is equal to $k_{\rm off}=0.2$~s$^{-1}$ \cite{J93}.  Since the polymerization rate at full speed is $k^{\rm p}_{+,{\rm max}}\simeq 300$~nt/s, the processivity of T7 DNA polymerase takes the value $l_{\rm max}\simeq 1500$~nt \cite{J93}.  Therefore, the processivity is large enough to justify the assumption of steady growth in order to obtain the properties of copolymerization.  

\begin{table}
\caption{\label{tab.T7-simul} Exo$^-$ T7 DNA polymerase at $20^{\circ}$C: The rate constants and other parameters used in the numerical simulations and the Markov-chain model.  The rate constants are from Refs.~\cite{PWJ91,WPJ91,J93}.  The other parameters are from the numerical simulations.}
\vspace{5mm}
\begin{center}
\begin{tabular}{|ccc|}
\hline
parameter & value & units \\
\hline
$k^{\rm p}_{+{\rm c}\vert{\rm c}}$ & $300$ & s$^{-1}$\\
$k^{\rm p}_{+{\rm i}\vert{\rm c}}$ & $0.03$ & s$^{-1}$\\
$k^{\rm p}_{+{\rm c}\vert{\rm i}}$ & $0.01$ & s$^{-1}$\\
$k^{\rm p}_{+{\rm i}\vert{\rm i}}$ & $0.01$ & s$^{-1}$\\
$K_{{\rm c}\vert{\rm c}}$ & $20$ & $\mu$M\\
$K_{{\rm i}\vert{\rm c}}$ & $6000$ & $\mu$M\\
$K_{{\rm c}\vert{\rm i}}$ & $84$ & $\mu$M\\
$K_{{\rm i}\vert{\rm i}}$ & $6000$ & $\mu$M\\
$K_{\rm P}$ & $200$ & mM\\
$[{\rm dNTP}]_{\rm eq}$ & $9.98\times 10^{-9}$ & M\\
$\eta_{\rm eq}$ & $2.41\times 10^{-3}$ & nt$^{-1}$\\
$D_{\rm eq}$ & $1.96\times 10^{-2}$ & nt$^{-1}$\\
$\eta_{\infty}$ & $1.04\times 10^{-6}$ & nt$^{-1}$\\
$D_{\infty}$ & $1.65\times 10^{-5}$ & nt$^{-1}$\\
$v_{\infty}$ & $288$ & nt/s\\
\hline
\end{tabular}
\end{center}
\end{table} 

\begin{table}
\caption{\label{tab.T7-B} Exo$^-$ T7 DNA polymerase at $20^{\circ}$C: The rate constants and other parameters of the Bernoulli-chain model.  The rate constants have been inferred Refs.~\cite{PWJ91,WPJ91,J93} and the other parameter have been calculated from theory.}
\vspace{5mm}
\begin{center}
\begin{tabular}{|ccc|}
\hline
parameter & value & units \\
\hline
$k^{\rm p}_{+{\rm c}}$ & $300$ & s$^{-1}$\\
$k^{\rm p}_{+{\rm i}}$ & $0.03$ & s$^{-1}$\\
$K_{\rm c}$ & $20$ & $\mu$M\\
$K_{\rm i}$ & $6000$ & $\mu$M\\
$K_{\rm P}$ & $200$ & mM\\
$[{\rm dNTP}]_{\rm eq,B}$ & $9.9\times 10^{-9}$ & M\\
$\eta_{\rm eq,B}$ & $9.9\times 10^{-3}$ & nt$^{-1}$\\
$D_{\rm eq,B}$ & $6.65\times 10^{-2}$ & nt$^{-1}$\\
$\eta_{\infty,{\rm B}}$ & $1\times10^{-6}$ & nt$^{-1}$\\
$D_{\infty,{\rm B}}$ & $1.59\times 10^{-5}$ & nt$^{-1}$\\
$v_{\infty,{\rm B}}$ & $297$ & nt/s\\
\hline
\end{tabular}
\end{center}
\end{table} 

\subsection{Numerical and theoretical results}

The kinetics is numerically simulated as a stochastic process by using Gillespie's algorithm~\cite{G76,G77}.  The details of this algorithm are given in Appendix~\ref{AppC}.  The concentrations of the four nucleotides are here supposed to be equal, which defines the nucleotide concentration~(\ref{dNTP}).  The template is taken as a Bernoulli chain of equal probabilities $\nu_1(n)=\frac{1}{4}$ for $n\in\{{\rm A},{\rm C},{\rm G},{\rm T}\}$.  For every value of dNTP concentration, the growth of $10^3$ chains each of length $10^6$ is numerically simulated and the different quantities of interest are obtained by statistical averages over this sample.  The results of numerical simulations are plotted as dots in the following figures, the quantities of the Markov-chain model as solid lines, and those of the Bernoulli-chain model as dashed lines.  These lines are calculated by solving the analytical equations given in Section~\ref{B-chain} for the Bernoulli-chain model, and in Section~\ref{M-chain} and Appendix~\ref{AppB} for the Markov-chain model.  Since the rates of T7 DNA polymerase are only known for correct and incorrect pairings without further distinction, Gillespie's algorithm actually simulates the Markov-chain model so that no difference is here expected between the properties of both (up to statistical errors).

\begin{figure}[h]
\centerline{\scalebox{0.55}{\includegraphics{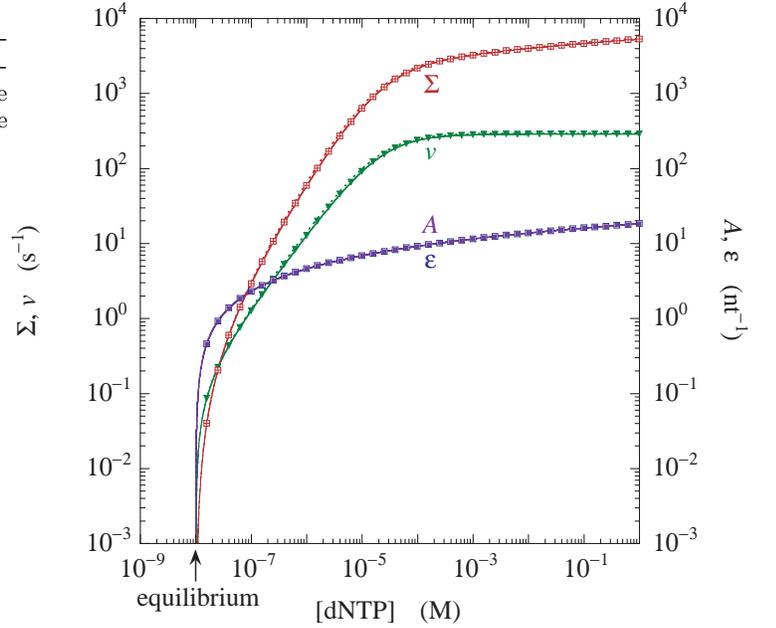}}}
\caption{Exo$^-$ T7 DNA polymerase: Entropy production $\Sigma$ (crossed squares), mean growth velocity $v$ (filled triangles), affinity $A$ (filled squares), and free-energy driving force $\epsilon$ (open squares) versus nucleotide concentration.  The dots are the results of numerical simulations, the solid lines of the Markov-chain model, and the dashed lines of the Bernoulli-chain model.}
\label{fig2}
\end{figure}

\begin{figure}[h]
\centerline{\scalebox{0.53}{\includegraphics{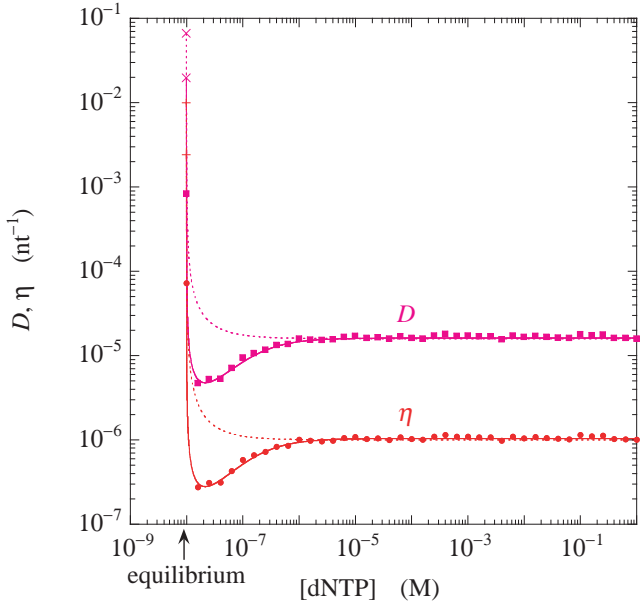}}}
\caption{Exo$^-$ T7 DNA polymerase: Conditional Shannon disorder per nucleotide $D$ (filled squares) and error probability $\eta$ (filled circles) versus nucleotide concentration.  The dots are the results of numerical simulations, the solid lines of the Markov-chain model, and the dashed lines of the Bernoulli-chain model.  The equilibrium values of the conditional Shannon disorder are shown as crosses and those of the error probability as pluses in both models.}
\label{fig3}
\end{figure}

Figure~\ref{fig2} shows the mean growth velocity~(\ref{velocity}), the entropy production~(\ref{entr-prod}), the affinity~(\ref{A}), and the free-energy driving force~(\ref{eps})-(\ref{eps_l}) as a function of the nucleotide concentration [dNTP].  The growth velocity behaves as expected by Eq.~(\ref{MM-velocity}) for a Michaelis-Menten kinetics.  Accordingly, the growth velocity, the entropy production, as well as the affinity are vanishing at the critical nucleotide concentration corresponding to the thermodynamic equilibrium.  The numerical value of $[{\rm dNTP}]_{\rm eq}$ given in Table~\ref{tab.T7-simul} corresponds to the theoretical value~(\ref{dNTP-eq-M}) of the Markov-chain model and is very close to the value~(\ref{dNTP-eq-B}) of the Bernoulli-chain model given in Table~\ref{tab.T7-B}.  At large values of dNTP concentration, the mean growth velocity reaches a plateau value of $288$~nt/s, which is in agreement with the values~(\ref{v-infty-B0}) and~(\ref{v-infty-M0}) of the Bernoulli- and Markov-chain models.  Besides, both the entropy production and the affinity are increasing logarithmically with the dNTP concentration, as described by Eqs.~(\ref{entr-prod-infty-B}), (\ref{A-infty-B}), (\ref{A-infty-M}), and~(\ref{entr-prod-infty-M}).  In Fig.~\ref{fig2}, the theoretical values of the different quantities for the Bernoulli-chain model are close to the numerical results obtained with Gillespie's algorithm, which here precisely simulates the Markov-chain model.

The corresponding error probability and conditional Shannon disorder per nucleotide are depicted in Fig.~\ref{fig3} versus dNTP concentration.  We see that the numerical results (dots) agree with the theoretical values (solid lines) of the Markov-chain model, but differences appear with respect to the Bernoulli-chain model (dashed lines) in the regime close to equilibrium.  We notice that, in every case, the conditional Shannon disorder is evaluated by Eq.~(\ref{D-estim}).  At full speed, the error probability takes the very small value $\eta_{\infty}\simeq 1.04\times 10^{-6}$~nt$^{-1}$, showing that T7 DNA polymerase has a high fidelity.  The accurate value given by Eq.~(\ref{eta-infty-M0}) is very close to its approximation~(\ref{eta-infty-M}), which here coincides with the value~(\ref{eta-infty-B}) of the Bernoulli-chain model given in Table~\ref{tab.T7-B}.  As the dNTP concentration is decreased towards the regime close to equilibrium, the error probability of the Markov-chain model slightly decreases to increase up to the value $\eta_{\rm eq}\simeq 2.41\times 10^{-3}$~nt$^{-1}$ well estimated by Eq.~(\ref{eta-eq-M}).  In the Bernoulli-chain model, the error probability monotonously increases to the even larger equilibrium value $\eta_{\rm eq,B}\simeq 9.9\times 10^{-3}$~nt$^{-1}$ given by Eqs.~(\ref{eta-eq-B0}) or~(\ref{eta-eq-B}).  We notice that the error probability is significantly larger at equilibrium than at full speed, $\eta_{\rm eq} \gg \eta_{\infty}$, for the T7 DNA polymerase.

\begin{figure}[h]
\centerline{\scalebox{0.55}{\includegraphics{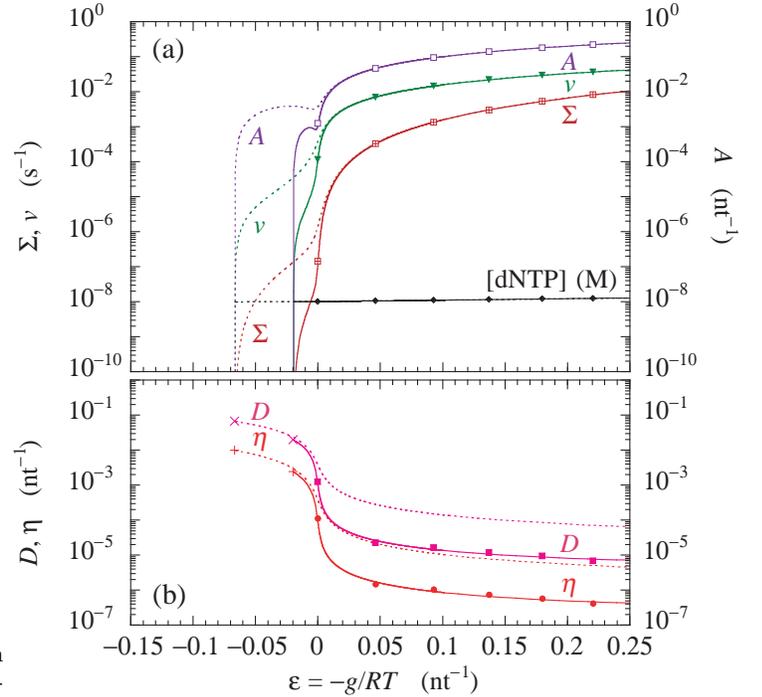}}}
\caption{Exo$^-$ T7 DNA polymerase: (a)~Affinity $A$ (open squares), mean growth velocity $v$ (filled triangles), entropy production $\Sigma$ (crossed squares), and nucleotide concentration [dNTP] (filled diamonds) versus the free-energy driving force $\epsilon$ in the regime close to equilibrium.  (b)~The corresponding conditional Shannon disorder $D$ (filled squares) and error probability $\eta$ (filled circles) versus the free-energy driving force $\epsilon$ in the same regime.  The equilibrium values of the conditional Shannon disorder are shown as crosses and those of the error probability as pluses in both models.  In (a) and (b), the dots are the results of numerical simulations, the solid lines of the Markov-chain model, and the dashed lines of the Bernoulli-chain model.}
\label{fig4}
\end{figure}

Figure~\ref{fig4} shows the different quantities as a function of the free-energy driving force $\epsilon=-g/(RT)$, which provides a magnification of the behavior in the regime close to equilibrium and a comparison with previous results~\cite{AG08}.  The plot reveals the crossover from the regime for $\epsilon>0$ where the growth is driven by free energy to the regime $-D_{\rm eq}<\epsilon\leq 0$ where the growth is driven by the entropic effect of disorder in the growing copy for both the Bernoulli- and Markov-chain models \cite{AG08}.  The growth velocity $v$, the entropy production $\Sigma$, and the affinity $A$ are vanishing at equilibrium when $\epsilon=-D_{\rm eq}$.  These quantities slowly increase in the regime of disorder-driven growth for $\epsilon<0$, and much more for $\epsilon >0$.  In contrast, the error probability $\eta$ and the conditional Shannon disorder per nucleotide $D$ decrease in the crossover as the free-energy driving force $\epsilon$ increases.  For this polymerase, the behavior is similar to the one already observed in Ref.~\cite{AG08}, although the equilibrium value of the conditional disorder is here smaller than in the Bernoulli-chain model adopted in Ref.~\cite{AG08} where the rates of the reversed reactions have been globally related to the free-energy driving force according to $W^{\rm p}_{-m:n}=W^{\rm p}_{+m:n}\exp(-\epsilon)$, which is a  simplification of the detailed kinetics.  However, we observe in Fig.~\ref{fig4} that the Markov-chain model has a smaller error probability at equilibrium than the Bernoulli-chain model.  Consequently, the corresponding conditional Shannon disorder is also smaller in the Markov-chain model by a factor $3.4$ with respect to the Bernoulli one, which explains that the solid lines remain closer to $\epsilon=0$ than the dashed lines in Fig.~\ref{fig4}.

We notice that the dissociation of the enzyme-DNA complex should limit the range of validity to growth velocities larger than the dissociation rate $v>k_{\rm off}=0.2$~s$^{-1}$, i.e., to concentrations $[{\rm dNTP}]>2.4\times10^{-8}$~M for the T7 DNA polymerase.  In this regard, the results about the regime close to the thermodynamic equilibrium may be more of theoretical interest for the steady growth regime than of experimental relevance.

\section{Human mitochondrial DNA polymerase}
\label{Hum-Pol}

\subsection{Phenomenology}

Human mitochondrial DNA polymerase $\gamma$ is responsible for the replication of mitochondrial genome coding for 13 proteins, 2 ribosomial RNAs, and 22 transfer RNAs in mitochondria \cite{JJ01a,LJ06}.  This polymerase is composed of two subunits: a catalytic protein of 140~kDa containing the polymerase and exonuclease domains and an accessory protein of 54~kDa \cite{JJ01a}.  Detailed experimental data have been obtained for an exonuclease-deficient mutant \cite{JJ01a,LJ06}.  The data used in the numerical simulations are given in Tables~\ref{tab.Hum-simul1} and~\ref{tab.Hum-simul2}.  Here, the polymerization rate and Michaelis-Menten dissociation constants are known for the sixteen possible pairings.  The rate constants of the reverse reactions are obtained by Eq.~(\ref{K_P-dfn}) using the same constant $K_{\rm P}=200$~mM as for the other polymerase.  Moreover, the pyrophosphate concentration is again fixed to the value $[{\rm P}]=10^{-4}$~M.

The processivity is also high for human mitochondrial DNA polymerase because the dissociation rate of the enzyme-DNA complex is here equal to $k_{\rm off}=0.02$~s$^{-1}$ while the maximal polymerization rate is $k^{\rm p}_{+,{\rm max}}\simeq 37$~nt/s, giving the value $l_{\rm max}\simeq 1850$~nt \cite{JJ01a}, which justifies the assumption of steady growth for this polymerase as well.

\begin{table}
\caption{\label{tab.Hum-simul1} Exo$^-$ human mitochondrial DNA polymerase $\gamma$ at $37^{\circ}$C: The polymerization rate constants and Michaelis-Menten dissociation constants used in the numerical simulations for a nucleotide attachment following a correct incorporation.  The data are from Ref.~\cite{LJ06}.}
\vspace{5mm}
\begin{center}
\begin{tabular}{|ccc|}
\hline
$m$:$n$ & $k^{\rm p}_{{+m\atop \ \, n}\vert{\rm c}}$ & $K_{{m\atop n}\vert{\rm c}}$ \\
pair &  s$^{-1}$ & $\mu$M \\
\hline
A:T & $45$ & $0.8$\\
A:G & $0.042$ & $250$\\
A:C & $0.1$ & $160$\\
A:A & $0.0036$ & $25$\\
C:T & $0.038$ & $360$\\
C:G & $43$ & $0.9$\\
C:C & $0.003$ & $140$\\
C:A & $0.1$ & $540$\\
G:T & $1.16$ & $70$\\
G:G & $0.066$ & $150$\\
G:C & $37$ & $0.8$\\
G:A & $0.1$ & $1000$\\
T:T & $0.013$ & $57$\\
T:G & $0.16$ & $200$\\
T:C & $0.012$ & $180$\\
T:A & $25$ & $0.6$\\
\hline
\end{tabular}
\end{center}
\end{table} 

\begin{table}
\caption{\label{tab.Hum-simul2} Exo$^-$ human mitochondrial DNA polymerase $\gamma$ at $37^{\circ}$C: Other rate constants from Ref.~\cite{JJ01a} used in the numerical simulations.  The parameters are from the numerical simulations.}
\vspace{5mm}
\begin{center}
\begin{tabular}{|ccc|}
\hline
parameter & value & units \\
\hline
$k^{\rm p}_{+{\rm c}\vert{\rm i}}$ & $0.52$ & s$^{-1}$\\
$k^{\rm p}_{+{\rm i}\vert{\rm i}}$ & $0.154$ & s$^{-1}$\\
$K_{{\rm c}\vert{\rm i}}$ & $404$ & $\mu$M\\
$K_{{\rm i}\vert{\rm i}}$ & $404$ & $\mu$M\\
$K_{\rm P}$ & $200$ & mM\\
$[{\rm dNTP}]_{\rm eq}$ & $3.87\times 10^{-10}$ & M\\
$\eta_{\rm eq}$ & $4.2\times 10^{-5}$ & nt$^{-1}$\\
$D_{\rm eq}$ & $5.1\times 10^{-4}$ & nt$^{-1}$\\
$\eta_{\infty}$ & $1.68\times 10^{-4}$ & nt$^{-1}$\\
$D_{\infty}$ & $1.8\times 10^{-3}$ & nt$^{-1}$\\
$v_{\infty}$ & $34$ & nt/s\\
\hline
\end{tabular}
\end{center}
\end{table} 

We notice in Table~\ref{tab.Hum-simul1} that the polymerization rate constants are much larger for correct than incorrect base pairing, while the Michaelis-Menten dissociation constants are smaller for correct than incorrect ones.  Moreover, the experimental data in Table~\ref{tab.Hum-simul2} show that the polymerization rates are significantly smaller after an incorrect incorporation, which again favors the Markov-chain model with respect to the Bernoulli one.

\subsection{Numerical and theoretical results}

Using the data of Tables~\ref{tab.Hum-simul1} and~\ref{tab.Hum-simul2}, the kinetics is here also numerically simulated as a stochastic process by using Gillespie's algorithm~\cite{G76,G77}.  See Appendix~\ref{AppC} for details.  Again, the equality~(\ref{dNTP}) of the four nucleotide concentrations is assumed and the template is taken as a Bernoulli chain of equal probabilities.  The statistics is performed with $10^3$ chains of length $10^6$ each.  Here, Gillespie's algorithm simulates the full kinetics with different rates for the sixteen nucleotide pairings even if the dNTP concentrations are equal.  In contrast to the situation in previous Section~\ref{T7-Pol}, we thus expect differences with respect to both Markov- and Bernoulli-chain models where only correct and incorrect pairings are distinguished.

The parameters of the Markov-chain and Bernoulli-chain models have been fitted to the results of the numerical simulations.  The corresponding parameter values are respectively given in Tables~\ref{tab.Hum-M} and~\ref{tab.Hum-B}. 

\begin{table}
\caption{\label{tab.Hum-M} Exo$^-$ human mitochondrial DNA polymerase $\gamma$ at $37^{\circ}$C: The rate constants and other parameters of the Markov-chain model fitted to the results of the numerical simulations.}
\vspace{5mm}
\begin{center}
\begin{tabular}{|ccc|}
\hline
parameter & value & units \\
\hline
$k^{\rm p}_{+{\rm c}\vert{\rm c}}$ & $37.3$ & s$^{-1}$\\
$k^{\rm p}_{+{\rm i}\vert{\rm c}}$ & $0.2628$ & s$^{-1}$\\
$k^{\rm p}_{+{\rm c}\vert{\rm i}}$ & $0.3$ & s$^{-1}$\\
$k^{\rm p}_{+{\rm i}\vert{\rm i}}$ & $0.01$ & s$^{-1}$\\
$K_{{\rm c}\vert{\rm c}}$ & $0.774$ & $\mu$M\\
$K_{{\rm i}\vert{\rm c}}$ & $107$ & $\mu$M\\
$K_{{\rm c}\vert{\rm i}}$ & $404$ & $\mu$M\\
$K_{{\rm i}\vert{\rm i}}$ & $404$ & $\mu$M\\
$K_{\rm P}$ & $200$ & mM\\
$[{\rm dNTP}]_{\rm eq,M}$ & $3.87\times 10^{-10}$ & M\\
$\eta_{\rm eq,M}$ & $4.2\times 10^{-5}$ & nt$^{-1}$\\
$D_{\rm eq,M}$ & $5.1\times 10^{-4}$ & nt$^{-1}$\\
$\eta_{\infty,{\rm M}}$ & $1.68\times10^{-4}$ & nt$^{-1}$\\
$D_{\infty,{\rm M}}$ & $1.8\times 10^{-3}$ & nt$^{-1}$\\
$v_{\infty,{\rm M}}$ & $34$ & nt/s\\
\hline
\end{tabular}
\end{center}
\end{table} 

\begin{table}
\caption{\label{tab.Hum-B} Exo$^-$ human mitochondrial DNA polymerase $\gamma$ at $37^{\circ}$C: The rate constants and other parameters of the Bernoulli-chain model fitted to the results of the numerical simulations.}
\vspace{5mm}
\begin{center}
\begin{tabular}{|ccc|}
\hline
parameter & value & units \\
\hline
$k^{\rm p}_{+{\rm c}}$ & $34.8$ & s$^{-1}$\\
$k^{\rm p}_{+{\rm i}}$ & $0.263$ & s$^{-1}$\\
$K_{\rm c}$ & $0.791$ & $\mu$M\\
$K_{\rm i}$ & $107$ & $\mu$M\\
$K_{\rm P}$ & $200$ & mM\\
$[{\rm dNTP}]_{\rm eq,B}$ & $3.87\times 10^{-10}$ & M\\
$\eta_{\rm eq,B}$ & $2.17\times 10^{-2}$ & nt$^{-1}$\\
$D_{\rm eq,B}$ & $1.29\times 10^{-1}$ & nt$^{-1}$\\
$\eta_{\infty,{\rm B}}$ & $1.68\times10^{-4}$ & nt$^{-1}$\\
$D_{\infty,{\rm B}}$ & $1.8\times 10^{-3}$ & nt$^{-1}$\\
$v_{\infty,{\rm B}}$ & $34$ & nt/s\\
\hline
\end{tabular}
\end{center}
\end{table} 

In the following figures, the results are depicted as dots for the numerical simulations, solid lines for the Markov-chain model, and dashed lines for the Bernoulli-chain model.  These lines are calculated thanks to the analytical methods given in Section~\ref{B-chain} for the Bernoulli-chain model, and in Section~\ref{M-chain} and Appendix~\ref{AppB} for the Markov-chain model.

\begin{figure}[h]
\centerline{\scalebox{0.55}{\includegraphics{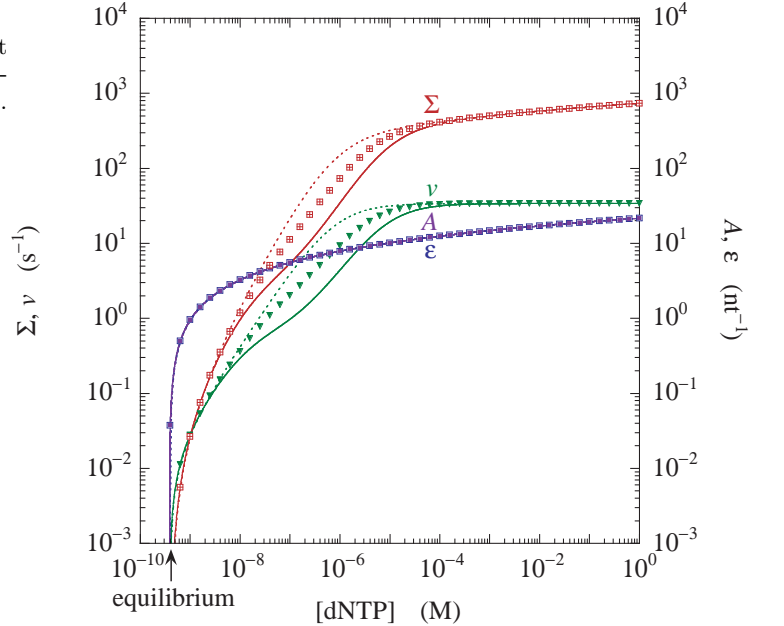}}}
\caption{Exo$^-$ human mitochondrial DNA polymerase: Entropy production $\Sigma$ (crossed squares), mean growth velocity $v$ (filled triangles), affinity $A$ (filled squares), and free-energy driving force $\epsilon$ (open squares) versus nucleotide concentration.  The dots are the results of numerical simulations, the solid lines of the Markov-chain model, and the dashed lines of the Bernoulli-chain model.}
\label{fig5}
\end{figure}

\begin{figure}[h]
\centerline{\scalebox{0.53}{\includegraphics{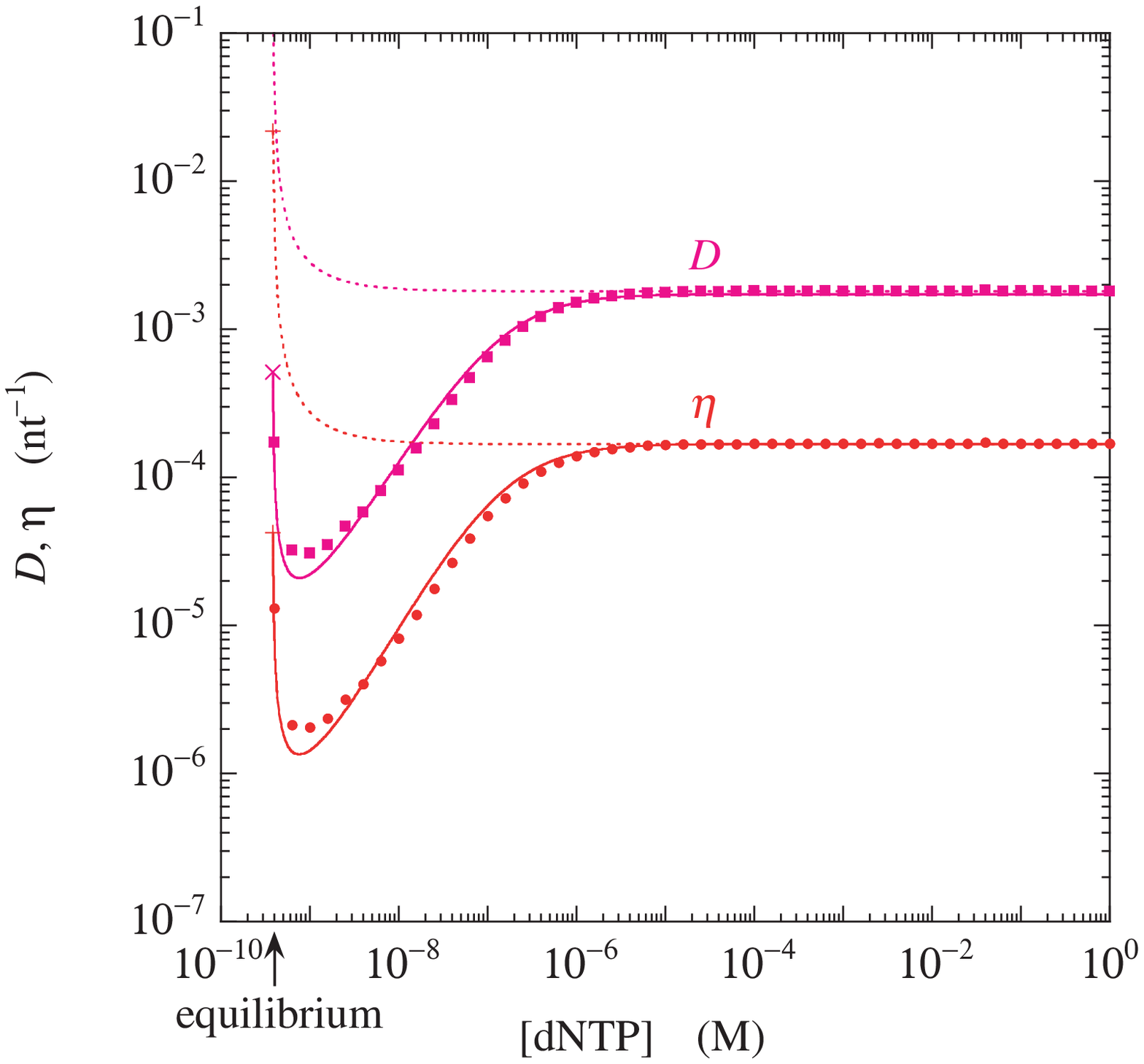}}}
\caption{Exo$^-$ human mitochondrial DNA polymerase: Conditional Shannon disorder per nucleotide $D$ (filled squares) and error probability $\eta$ (filled circles) versus nucleotide concentration.  The dots are the results of numerical simulations, the solid lines of the Markov-chain model, and the dashed lines of the Bernoulli-chain model.  The equilibrium values of the conditional Shannon disorder are shown as crosses and those of the error probability as pluses in both models.}
\label{fig6}
\end{figure}

For the human mitochondrial DNA polymerase, Fig.~\ref{fig5} depicts the mean growth velocity~(\ref{velocity}), the entropy production~(\ref{entr-prod}), the affinity~(\ref{A}), and the free-energy driving force~(\ref{eps})-(\ref{eps_l}) as a function of the nucleotide concentration [dNTP].  As before, the velocity, the entropy production, and the affinity vanish at the equilibrium concentration $[{\rm dNTP}]_{\rm eq}$, which is well approximated in both models by Eqs.~(\ref{dNTP-eq-B}) and~(\ref{dNTP-eq-M}), as seen in Tables~\ref{tab.Hum-simul2}-\ref{tab.Hum-B}.  By increasing dNTP concentration, the polymerase reaches its full speed regime where the mean growth velocity culminates at the value $v_{\infty}\simeq 34$~nt/s for this slower polymerase than the T7 DNA polymerase.  This maximum velocity is also well approximated in both models by Eqs.~(\ref{v-infty-B0}) and~(\ref{v-infty-M0}).  The entropy production and the affinity continue to increase with the dNTP concentration, as predicted by Eqs.~(\ref{entr-prod-infty-B}), (\ref{A-infty-B}), (\ref{A-infty-M}), and~(\ref{entr-prod-infty-M}).  However, discrepancies appear in the intermediate regime between the numerical simulations and both the Markov- and Bernoulli-chain models.  The reason is that the numerically simulated system is here richer than both the Markov- and Bernoulli-chain models, which have too few parameters to reproduce the results of the full simulation. Nevertheless, the general behavior is qualitatively reproduced by both models for the quantities depicted in Fig.~\ref{fig5}.

Figure~\ref{fig6} shows the error probability and conditional Shannon disorder per nucleotide versus dNTP concentration, corresponding to the conditions of Fig.~\ref{fig5}.  We only discuss the behavior of the error probability because the conditional Shannon disorder is evaluated by Eq.~(\ref{D-estim}) in every case.  At full speed, the results are in agreement between the numerical simulations (dots), the Markov-chain model (solid lines), and the Bernoulli-chain model (dashed lines).  In this regime, the error probability takes the value $\eta_{\infty}\simeq 1.68\times 10^{-4}$~nt$^{-1}$, confirming that the human mitochondrial DNA polymerase has a lower fidelity than T7 DNA polymerase.  The full speed value of the error probability is very well approximated by Eqs.~(\ref{eta-infty-B0}) and~(\ref{eta-infty-M0}).  The approximation~(\ref{eta-infty-M}) gives the value $\eta_{\infty,{\rm M}}\simeq 1.53\times 10^{-4}$, which is still a close estimation.

However, differences appear at low values of dNTP concentration where equilibrium is approached.  As the dNTP concentration is decreased, the error probability of the numerical simulation and Markov-chain model decreases more significantly than in Fig.~\ref{fig3} for T7 DNA polymerase, before increasing slightly very close to equilibrium.  Instead the error probability of the Bernoulli-chain model only increases, showing the shortcoming of this model close to equilibrium.  At variance with respect to the case of T7 DNA polymerase, the equilibrium error probability of the simulation and the Markov-chain model is lower than at full speed, $\eta_{\rm eq}\simeq 4.2\times 10^{-5}  < \eta_{\infty}\simeq 1.68\times 10^{-4}$, while the Bernoulli-chain model gives a much larger equilibrium error probability $\eta_{\rm eq,B}\simeq 2.17\times 10^{-2}$.  The reason is that the equilibrium error probability has an extra factor smaller than unity in Eq.~(\ref{eta-eq-M}) for the Markov-chain model, than in Eq.~(\ref{eta-eq-B}) for the Bernoulli-chain model.

\begin{figure}[h]
\centerline{\scalebox{0.55}{\includegraphics{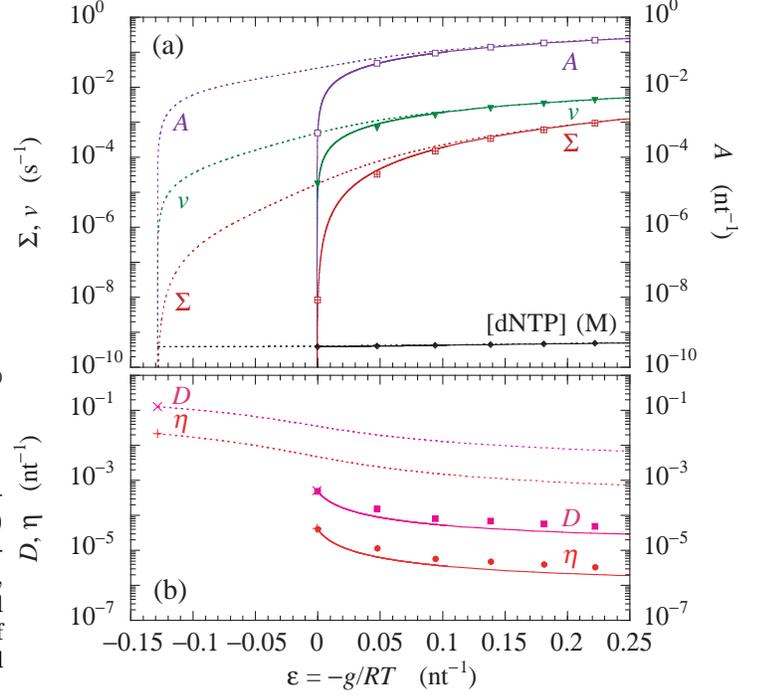}}}
\caption{Exo$^-$ human mitochondrial DNA polymerase: (a)~Affinity $A$ (open squares), mean growth velocity $v$ (filled triangles), entropy production $\Sigma$ (crossed squares), and nucleotide concentration [dNTP] (filled diamonds) versus the free-energy driving force $\epsilon$ in the regime close to equilibrium.  (b)~The corresponding conditional Shannon disorder $D$ (filled squares) and error probability $\eta$ (filled circles) versus the free-energy driving force $\epsilon$ in the same regime.  The equilibrium values of the conditional Shannon disorder are shown as crosses and those of the error probability as pluses in both models.  In (a) and (b), the dots are the results of numerical simulations, the solid lines of the Markov-chain model, and the dashed lines of the Bernoulli-chain model.}
\label{fig7}
\end{figure}

In Fig.~\ref{fig7}, the different quantities are shown as a function of the free-energy driving force $\epsilon=-g/(RT)$ in the regime close to equilibrium.  The crossover is observed from the regime of growth driven by free energy if $\epsilon>0$ to the regime of disorder-driven growth if $-D_{\rm eq}<\epsilon\leq 0$ \cite{AG08}.  The crossover is clear for the Bernoulli-chain model because the conditional Shannon disorder takes the large value $D_{\rm eq,B}\simeq 0.129$ in this model.  Instead, the disorder is much smaller in the Markov-chain model where $D_{\rm eq,M}\simeq 5.1\times 10^{-4}$, which explains that it is not visible with the scale used for the driving force~$\epsilon$ in Fig.~\ref{fig7}. In both models, the growth velocity $v$, the entropy production $\Sigma$, and the affinity $A$ vanish at their respective equilibrium value $\epsilon=-D_{\rm eq}$, while the error probability $\eta$ and the conditional Shannon disorder per nucleotide $D$ increase as equilibrium is approached.  The behavior is reminiscent of the one observed in Ref.~\cite{AG08}, although the equilibrium conditional disorder is here smaller especially in the Markov-chain model.

Here also, the dissociation of the enzyme-DNA complex should limit the experimental relevance of the study to the range where the growth velocities are larger than the dissociation rate $v>k_{\rm off}=0.02$~s$^{-1}$, i.e., to concentrations $[{\rm dNTP}]>8.2\times10^{-10}$~M for the human mitochondrial DNA polymerase.

\section{Discussion}
\label{Discussion}

In the present paper, the kinetic theory of exonuclease-deficient DNA polymerases has been developed in order to determine the speed, fidelity, and thermodynamics of DNA replication in terms of the biochemical rate constants, the concentrations of nucleotides and other substances, and the template sequence.  For this purpose, recent theoretical work has been used on the growth and thermodynamics of Bernoulli and Markov chains \cite{AG08,GA14,AG09}.  

Already without exonuclease proofreading, the kinetics of DNA polymerases is of great importance for understanding DNA replication and many experiments are specifically devoted to these enzymes.  Indeed, the discrimination between correct and incorrect nucleotides may already be quite efficient without dedicated proofreading mechanisms.  By explicitly taking into account the dependence of the rates on the concentrations of nucleotides and pyrophosphate, the theory provides direct comparison with experimental observations \cite{PWJ91,WPJ91,DPJ91,J93,TJ06,JJ01a,JJ01b,LNKC01,LJ06,EG91,KB00,SUKOOBWFWMG08,RBT08,ZBNS09,DJPW10,BBT12}.  In particular, the theory explains the Michaelis-Menten dependence of the mean growth velocity on the nucleotide concentration, which is a basic feature of enzymatic kinetics.

In the present paper, a systematic comparison is carried out between the Bernoulli- and Markov-chain models.  Until now, theoretical work has mainly used Bernoulli-chain models.  However, experimental observations have revealed that the rates of DNA polymerases depend not only on the pairing and polymerization of a new nucleotide, but also on the previously incorporated nucleotide \cite{WPJ91,J93,JJ01a}.  The reason is that, by their structure, the DNA polymerases have a mechanical interaction with DNA allowing their dynamics to depend on a few base pairs in the growing DNA and to be sensitive to possible mismatches caused by previously formed base pairs.  In this respect, a key role is played by conformational changes in DNA polymerases during elongation \cite{J93}.  These essential aspects imply that the copy growing on a Bernoullian template is a Markov chain, instead of a Bernoulli chain itself.

Results are obtained for the error probability, the thermodynamics of DNA replication, and their consequences on the evolution of sequences from generation to generation.

An important point is that the mutual information characterizing replication fidelity takes a value very close to the overall disorder if the error probability is low enough, as shown by Eq.~(\ref{I-estim}).  Therefore, the contribution of the conditional disorder~(\ref{Dc=D-I}) to the thermodynamic entropy production remains small to the extent that the kinetics of replication has a high fidelity, although the overall disorder takes larger values close to $\ln 4$ since it characterizes instead the static structure of aperiodic DNA sequences.  As shown with Eq.~(\ref{Dw-Da-N}), the replication tends to increase the overall disorder between the template and the copy without the further DNA mismatch repair mechanism.  Since the error probability is very small $\eta\ll 1$, the overall disorder slowly evolves between generations close to its maximal value $D(\omega)\simeq D(\alpha)\simeq \ln 4$.  

Furthermore, analytical expressions are deduced for the error probability.  In the full speed regime, the  error probability in the Bernoulli- and Markov-chain models can be expressed as
\bea
\eta_{\infty,{\rm B}} &\simeq& 3 \, \frac{k^{\rm p}_{+{\rm i}} \, K_{\rm c}}{k^{\rm p}_{+{\rm c}} \, K_{\rm i}} \, , \label{eta_infty_B}\\
\eta_{\infty,{\rm M}} &\simeq& 3 \, \frac{k^{\rm p}_{+{\rm i}\vert{\rm c}} \, K_{{\rm c}\vert{\rm c}}}{k^{\rm p}_{+{\rm c}\vert{\rm c}} \, K_{{\rm i}\vert{\rm c}}} \, , \label{eta_infty_M}
\eea
showing that fidelity is essentially controlled by the discrimination between correct and incorrect pairings after correct incorporation.  The inverse of the error probability~(\ref{eta_infty_B}) is known to characterize the fidelity of DNA polymerases \cite{BBT12}.  In particular, this quantity has been evaluated by the theoretical computation of the free-energy landscape along the conformational changes and the reaction pathway of DNA polymerases \cite{FGW05,XGBWW08,ROW10}.  The formula~(\ref{eta_infty_M}) generalizes this result to the Markov-chain case.  If the kinetic constants are experimentally measured for every possible nucleotide pairings, the present theory also shows that the error probability can be approximatively evaluated according to
\be
\eta_{\infty,{\rm M}} \simeq 3 \left\langle\frac{K_{{\rm c}\vert{\rm c}}}{k^{\rm p}_{+{\rm c}\vert{\rm c}}}\right\rangle \left\langle\frac{k^{\rm p}_{+{\rm i}\vert{\rm c}}}{K_{{\rm i}\vert{\rm c}}}\right\rangle \, , \label{eta_infty_M_av}
\ee
in terms of separate averages $\langle\cdot\rangle$ for the corresponding ratios of correct and incorrect pairings.

For the exo$^-$ DNA polymerase of T7 viruses, the expression~(\ref{eta_infty_M}) applied to the model of Section~\ref{T7-Pol} gives the error probability $\eta_{\infty}\simeq 10^{-6}$ in agreement with the range of experimental values $3\times 10^{-7}$-$7\times 10^{-6}$ reported in the literature \cite{WPJ91,JJ01a}.  For the exo$^-$ DNA polymerase of human mitochondria, which has a lower fidelity than the one for T7 viruses, the error probability takes the larger value $\eta_{\infty}\simeq 1.68\times 10^{-4}$ also in agreement with known experimental values of about $10^{-4}$ \cite{LNKC01}.

Results are also obtained for the error probability at low dNTP concentration in the regime close to  thermodynamic equilibrium.  In this regime, the present analysis reveals differences between the Bernoulli- and Markov-chain models.  A dip is observed in Figs.~\ref{fig3} and~\ref{fig6} for the error probability in the Markov-chain model, which is not the feature of the Bernoulli one.  Equilibrium happens at a critical dNTP concentration, which can be calculated in both the Bernoulli- and Markov-chain models, giving comparable values.  However, the error probability at equilibrium is much different in both models and it varies significantly depending on the parameter set of the polymerase.  Consistently, the equilibrium error probability only depends on the Michaelis-Menten constants associated with quasi-equilibria resulting from detailed balance between opposite transitions.  Under the assumption that dissociation is lower for correct than incorrect base pairs, the equilibrium error probability is evaluated as
\be
\eta_{\rm eq,B} \simeq 3 \, \frac{K_{\rm c}}{K_{\rm i}} 
\ee
in the Bernoulli-chain model, but as
\be
\eta_{\rm eq,M} \simeq 3 \, \frac{K_{{\rm c}\vert{\rm c}}^2}{K_{{\rm c}\vert{\rm i}} \, K_{{\rm i}\vert{\rm c}}} 
\ee
in the Markov-chain model.  Therefore, the equilibrium error probability can be much smaller in the Markov- than the Bernoulli-chain model.  Moreover, the error probability may be smaller at equilibrium than at full speed in the Markov-chain model, as observed in Figs.~\ref{fig6} and~\ref{fig7} for the parameter set of the exo$^-$ human mitochondrial DNA polymerase.  The present work thus shows that the behavior of exo$^-$ DNA polymerases in the regime close to equilibrium is very much sensitive to the dependence of kinetics on the previously incorporated nucleotide.  The difference could be observed experimentally by studying how error probability varies with nucleotide concentration.  We may expect that the computational approach could also bring more knowledge about the thermodynamics of DNA polymerases in the future.  

\begin{figure}[h]
\centerline{\scalebox{0.55}{\includegraphics{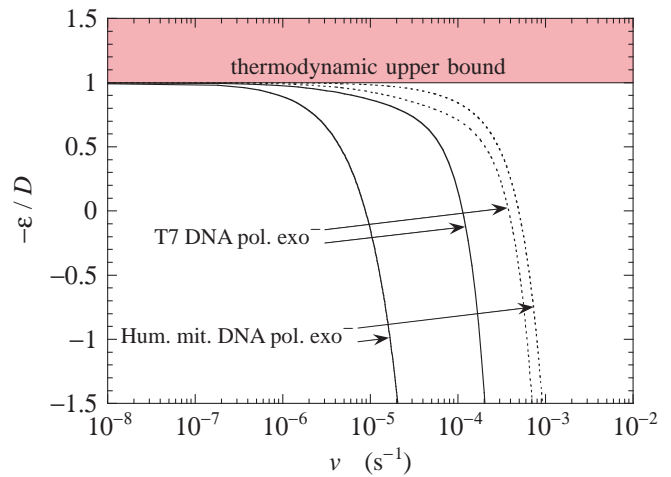}}}
\caption{Thermodynamic upper bound on minus the ratio of the free-energy driving force $\epsilon$ to the conditional Shannon disorder per nucleotide $D=D(\omega\vert\alpha)$ of the copy $\omega$ with respect to the template $\alpha$ versus the mean growth velocity $v$ for the T7 and human mitochondrial DNA polymerases in the Bernoulli- (dashed lines) and Markov-chain (solid lines) models.}
\label{fig8}
\end{figure}

In any case, the present theory predicts a thermodynamic upper bound on minus the ratio of the free-energy driving force $\epsilon$ to the conditional Shannon disorder per nucleotide $D(\omega\vert\alpha)$ in the growth regime where the velocity is positive $v>0$.  Indeed, the entropy production~(\ref{entr-prod}) is always non-negative according to the second law of thermodynamics.  If $v>0$, the affinity~(\ref{A}) should thus also be non-negative, leading to the thermodynamic inequality
\be
-\frac{\epsilon}{D(\omega\vert\alpha)} \leq 1 \, ,
\label{-e/D}
\ee
meaning that it is not possible to extract more free energy than provided by the conditional sequence disorder of the copy with respect to the template.  Figure~\ref{fig8} shows that this upper bound is indeed satisfied for both exo$^-$ DNA polymerases in the Bernoulli- and Markov-chain models.  The ratio~(\ref{-e/D}) reaches its maximal value equal to unity as thermodynamic equilibrium is approached when the growth velocity is vanishing.

The practical observation of this prediction requires that the conditional Shannon disorder and thus the error probability take large enough values.  This is the case for low-fidelity polymerases because their error probability can be as high as $0.1$-$0.5$ \cite{MBMHK00,FWR02,W-14}.  Besides, the fidelity of DNA polymerase is known to be reduced by the presence of manganese ions Mn$^{2+}$ in the surrounding solution, which is called manganese mutagenesis \cite{GKGB83,EDS84,BML85,TR89}.  Under such circumstances, the contribution of the conditional disorder to the thermodynamic entropy production should thus be larger, in particular, close to equilibrium, which could allow the experimental observation of the crossover from the regime of growth driven by the entropic effect of disorder to the one driven by free energy \cite{AG08}.  The dependence of these properties on an external force can also be envisaged \cite{WSYKB00,MBC00}.  Similar considerations apply to RNA polymerases.

In the companion paper~\cite{paperII}, the goal will be to extend the analysis to DNA polymerases with exonuclease proofreading.  For this purpose, the dependence of the rates on the previously incorporated nucleotide and the analytical methods developed for Markov-chain growth processes will turn out to play a crucial role.

\begin{acknowledgments}
The author is grateful to D. Andrieux, D. Bensimon, J. England, D. Lacoste, Y. Rondelez, and S.~A.~Rice for helpful discussions, remarks, and support during the elaboration of this work.
This research is financially supported by the Universit\'e Libre de Bruxelles, the FNRS-F.R.S., and the Belgian Federal Government under the Interuniversity Attraction Pole project P7/18 ``DYGEST".
\end{acknowledgments}

\begin{widetext}

\appendix

\section{Equations for kinetics and thermodynamics}
\label{AppA}

\subsection{Kinetics}

For the reaction network depicted in Fig.~\ref{fig1}, the kinetic equations ruling the time evolution of the probabilities~(\ref{probabilities}) are given by
\bea
\frac{d}{dt}\, {\cal P}_t\left(m_1 \cdots m_l \qquad\quad\ \atop n_1\, \cdots \, n_l \, n_{l+1}\cdots\right) &=& 
k^{\rm p}_{+m_l m_{l-1}\atop \ \, n_l \, n_{l-1}}
\, {\cal P}_t\left(m_1 \cdots m_l{\rm P} \qquad\ \ \atop n_1\, \cdots \, n_l \, n_{l+1}\cdots\right)
+\sum_{m_{l+1}} k_{-m_{l+1} m_l\atop \ \, n_{l+1} \, n_l} 
\, {\cal P}_t\left(m_1 \cdots m_l m_{l+1}{\rm P} \qquad\ \ \atop n_1\, \cdots \, n_l \, n_{l+1}\, n_{l+2}\cdots\right)
\nonumber\\
&-& \left( k^{\rm p}_{-m_l m_{l-1}\atop \ \, n_l \, n_{l-1}}[{\rm P}]
+\sum_{m_{l+1}} k_{+m_{l+1} m_l\atop \ \, n_{l+1} \, n_l}[m_{l+1}{\rm P}]\right)  
{\cal P}_t\left(m_1 \cdots m_l \qquad\quad \ \atop n_1\, \cdots \, n_l \, n_{l+1}\cdots\right) 
\label{kin_eq_1}
\eea
and
\bea
\frac{d}{dt}\, {\cal P}_t\left(m_1 \cdots m_l m_{l+1}{\rm P} \qquad\ \ \atop n_1\, \cdots \, n_l \, n_{l+1}\, n_{l+2}\cdots\right) &=& 
k_{+m_{l+1} m_l\atop \ \, n_{l+1} \, n_l}[m_{l+1}{\rm P}]
\, {\cal P}_t\left(m_1 \cdots m_l \qquad\quad\ \atop n_1\, \cdots \, n_l \, n_{l+1}\cdots\right)
+k^{\rm p}_{-m_{l+1} m_l\atop \ \, n_{l+1} \, n_l}[{\rm P}]
\, {\cal P}_t\left(m_1 \cdots m_l m_{l+1} \qquad\quad\ \atop n_1\, \cdots \, n_l \, n_{l+1}\, n_{l+2}\cdots\right)
\nonumber\\
&-& \left( k_{-m_{l+1} m_l\atop \ \, n_{l+1} \, n_l} + k^{\rm p}_{+m_{l+1} m_l\atop \ \, n_{l+1} \, n_l}\right)  
{\cal P}_t\left(m_1 \cdots m_l m_{l+1}{\rm P} \qquad\ \ \atop n_1\, \cdots \, n_l \, n_{l+1}\, n_{l+2}\cdots\right)
\label{kin_eq_2}
\eea
in terms of the rates~(\ref{nb_rate})-(\ref{depol_rate}) for $l=1,2,3,...$.  In Eq.~(\ref{kin_eq_1}) for the probability of a copy ending with a monophosphate group, the gain terms are due to polymerization by pyrophosphate release and to nucleotide dissociation, and the loss terms to depolymerization by pyrophosphorolysis and nucleotide binding.  In Eq.~(\ref{kin_eq_2}) for the probability of a copy ending with a triphosphate group, the gain terms are due to nucleotide binding and depolymerization, and the loss terms to nucleotide dissociation and polymerization.  For $l=1$ in Eq.~(\ref{kin_eq_1}) and $l=0$ in Eq.~(\ref{kin_eq_2}), the symbols $m_0$ and $n_0$ stand for the empty set: $m_0=n_0=\emptyset$.  For $l=0$, Eq.~(\ref{kin_eq_1})  should be replaced by
\be
\frac{d}{dt}\, {\cal P}_t\left(\emptyset\qquad\quad \ \atop n_1\, n_2 \, \cdots\right) =
\sum_{m_1} k_{-m_1 \emptyset \atop \ \ n_1\, \emptyset} 
\, {\cal P}_t\left(m_1{\rm P} \qquad\ \ \atop n_1\, n_2 \, \cdots\right)
-\sum_{m_1} k_{+m_1\emptyset\atop \ \ n_1\, \emptyset}[m_1{\rm P}]\, 
{\cal P}_t\left(\emptyset\qquad\quad \ \atop n_1\, n_2\cdots\right) \, ,
\label{kin_eq_0}
\ee
ruling the probability of the lone template bounded to the enzyme.
For $l=0$ and $l=1$, these kinetic equations describe the initiation of the copolymerization process.  We notice that the initiation steps become negligible as $l\to\infty$ in the regime of steady growth, which is here investigated. The equations (\ref{kin_eq_1})-(\ref{kin_eq_0}) preserve the total probability:
\be
\sum_l\sum_{m_1\cdots m_l}\left[{\cal P}_t\left(m_1 \cdots m_l \qquad\quad\ \atop n_1\, \cdots \, n_l \, n_{l+1}\cdots\right) 
+\sum_{m_{l+1}}
 {\cal P}_t\left(m_1 \cdots m_l m_{l+1}{\rm P}\qquad\ \ \atop n_1\, \cdots \, n_l \, n_{l+1}\, n_{l+2}\cdots\right)\right] =1 \, .
\ee

As explained in Subsection~\ref{MM_kin}, the quasi-equilibrium between nucleotide binding and dissociation resulting from the assumption~(\ref{MM-hyp}) implies that the kinetic equations (\ref{kin_eq_1})-(\ref{kin_eq_2}) reduce to a Michaelis-Menten kinetics for the following sum
\be
P_t\left(m_1 \cdots m_l \qquad\quad\ \atop n_1\, \cdots \, n_l \, n_{l+1}\cdots\right)
\equiv
{\cal P}_t\left(m_1 \cdots m_l \qquad\quad\ \atop n_1\, \cdots \, n_l \, n_{l+1}\cdots\right) 
+\sum_{m_{l+1}}
 {\cal P}_t\left(m_1 \cdots m_l m_{l+1}{\rm P} \qquad\ \ \atop n_1\, \cdots \, n_l \, n_{l+1}\, n_{l+2}\cdots\right)
\label{prob_sum}
\ee
of the probabilities~(\ref{probabilities}).  The time evolution of the probability~(\ref{prob_sum}) is ruled by the following kinetic equation:
\bea
\frac{d}{dt}\, P_t\left(m_1 \cdots m_l \qquad\quad\ \atop n_1\, \cdots \, n_l \, n_{l+1}\cdots\right) &=& 
W_{+m_l m_{l-1}\atop \ \, n_l \, n_{l-1}} 
\, P_t\left(m_1 \cdots m_{l-1} \qquad\ \ \atop n_1\, \cdots \, n_{l-1} \, n_l\cdots\right)
+\sum_{m_{l+1}} W_{\quad\, -m_{l+1} m_l\atop n_{l+2}\, n_{l+1} \, n_l}
\, P_t\left(m_1 \cdots m_l m_{l+1} \qquad\ \ \atop n_1\, \cdots \, n_l \, n_{l+1}\, n_{l+2}\cdots\right)
\nonumber\\
&-& \left( W_{\quad\, -m_l m_{l-1}\atop n_{l+1}\, n_l \, n_{l-1}}
+\sum_{m_{l+1}} W_{+m_{l+1} m_l\atop \ \, n_{l+1}\, n_l}\right)  
P_t\left(m_1 \cdots m_l \qquad\quad \ \atop n_1\, \cdots \, n_l \, n_{l+1}\cdots\right) 
\label{kin_eq}
\eea
with the rates
\be
W_{+m_{l+1} m_l\atop \ \, n_{l+1}\, n_l}= W^{\rm p}_{+m_{l+1} m_l\atop \ \, n_{l+1}\, n_l}
\qquad \mbox{and} \qquad
W_{\quad\, -m_l m_{l-1}\atop n_{l+1}\, n_l \, n_{l-1}}=W^{\rm p}_{\quad\, -m_l m_{l-1}\atop n_{l+1}\, n_l \, n_{l-1}}
\label{rates-A}
\ee
given by Eqs.~(\ref{Wp+}) and~(\ref{Wp-}) with the denominator~(\ref{denom}).  Again the total probability 
\be
\sum_l\sum_{m_1\cdots m_l}P_t\left(m_1 \cdots m_l \qquad\quad\ \atop n_1\, \cdots \, n_l \, n_{l+1}\cdots\right)=1
\ee
is preserved by the kinetic equations~(\ref{kin_eq}).

\subsection{Thermodynamics}

The connection with thermodynamics is established by noticing that the ratio of the rates for opposite transitions is related by
\be
\frac{W^{\rm p}_{+m_l m_{l-1}\atop \ \, n_l\, n_{l-1}}}
{W^{\rm p}_{\quad\, -m_l m_{l-1}\atop n_{l+1}\, n_l \, n_{l-1}}}=\exp\bigg[\beta \underbrace{G\left(m_1 \cdots m_{l-1} \qquad\ \ \atop n_1\, \cdots \, n_{l-1} \, n_l\cdots\right)}_{G_{l-1}(\omega\vert\alpha)}-\beta \underbrace{G\left(m_1 \cdots m_l \qquad\quad\ \ \atop n_1\, \cdots \, n_l \, n_{l+1}\cdots\right)}_{G_l(\omega\vert\alpha)}\bigg]
\label{W_ratio}
\ee
with
$\beta=(RT)^{-1}$ to the difference of free enthalpies of the two coarse-grained states, between which the transitions happen.  The entropy production is given by \cite{P55,S76,N79,LVN84,JQQ04,G04}
\bea
\frac{1}{R}\frac{d_{\rm i}S}{dt} &=& \sum_l \sum_{m_1\cdots m_l}
\left[W^{\rm p}_{+m_l m_{l-1}\atop \ \, n_l \, n_{l-1}} 
\, P_t\left(m_1 \cdots m_{l-1} \qquad\ \ \atop n_1\, \cdots \, n_{l-1} \, n_l\cdots\right)
-W^{\rm p}_{\quad\, -m_l m_{l-1}\atop n_{l+1}\, n_l \, n_{l-1}}
\, P_t\left(m_1 \cdots m_l \qquad\quad\ \ \atop n_1\, \cdots \, n_l \, n_{l+1}\cdots\right)\right]
\nonumber\\
&&\qquad\qquad\qquad\qquad\qquad\qquad
\times \ln\frac{W^{\rm p}_{+m_l m_{l-1}\atop \ \, n_l \, n_{l-1}} 
\, P_t\left(m_1 \cdots m_{l-1} \quad\ \ \ \atop n_1\, \cdots \, n_{l-1} \, n_l\cdots\right)}
{W^{\rm p}_{\quad\, -m_l m_{l-1}\atop n_{l+1}\, n_l \, n_{l-1}}
\, P_t\left(m_1 \cdots m_l \qquad\ \ \atop n_1\, \cdots \, n_l \, n_{l+1}\cdots\right)} \geq 0 \, .
\label{entr-A}
\eea
In the regime of steady growth, this expression becomes Eq.~(\ref{entr-prod}) in terms of the mean growth velocity~(\ref{velocity}), the affinity~(\ref{A}), the free-energy driving force
\be
\epsilon = - \frac{g}{RT} = \lim_{l\to\infty} -\frac{1}{l} \sum_{\alpha,\omega} \nu_l(\alpha)\, \mu_l(\omega\vert\alpha) \, \beta G_l(\omega\vert\alpha) \, ,
\label{eps-A}
\ee
which is equivalent to Eqs.~(\ref{eps})-(\ref{eps_l}) because of Eq.~(\ref{W_ratio}), and the conditional Shannon disorder per nucleotide~(\ref{Dc}) \cite{AG08,AG09}.

\section{Solving the Markov-chain model}
\label{AppB}

\subsection{Solution of the kinetic equations}

In the regime of steady growth, the kinetic equations~(\ref{kin_eq}) can be solved analytically in the form given by Eq.~(\ref{prob2}) with the factorization~(\ref{Markov_chain}) of a Markov chain running from the growing tip $m_l$ of the copy back to the first nucleotide $m_1$ in terms of the conditional and tip probabilities~(\ref{M_cond_prob}) and~(\ref{M_tip_prob}).  The analytical method has been presented in Ref.~\cite{GA14}.  In order to solve the problem, the partial velocities~(\ref{M_part_velocities}) are first calculated by the following self-consistent equations:
\bea
v_{\rm c} &=& \frac{W_{+{\rm c}\vert{\rm c}}\, v_{\rm c}}{W_{-{\rm c}\vert{\rm c}} + v_{\rm c}} + 3\, \frac{W_{+{\rm i}\vert{\rm c}}\, v_{\rm i}}{W_{-{\rm i}\vert{\rm c}} + v_{\rm i}} \, , \label{v_c-A}\\
v_{\rm i} &=& \frac{W_{+{\rm c}\vert{\rm i}}\, v_{\rm c}}{W_{-{\rm c}\vert{\rm i}} + v_{\rm c}} + 3\, \frac{W_{+{\rm i}\vert{\rm i}}\, v_{\rm i}}{W_{-{\rm i}\vert{\rm i}} + v_{\rm i}} \, , \label{v_i-A}
\eea
in terms of the rates~(\ref{rates-A}) given by Eqs.~(\ref{Wp+})-(\ref{denom}). These self-consistent equations can be solved by numerical iterations in order to get the partial velocities $v_{\rm c}$ and $v_{\rm i}$, starting from some positive initial values.  Thereafter, the tip probabilities~(\ref{M_tip_prob}) are calculated with
\bea
\mu({\rm c}) &=& \frac{W_{+{\rm c}\vert{\rm c}}}{W_{-{\rm c}\vert{\rm c}} + v_{\rm c}}\, \mu({\rm c}) + 3\, \frac{W_{+{\rm c}\vert{\rm i}}}{W_{-{\rm c}\vert{\rm i}} + v_{\rm c}}\, \mu({\rm i}) \, , \label{mu_c-A}\\
\mu({\rm i}) &=& \frac{W_{+{\rm i}\vert{\rm c}}}{W_{-{\rm i}\vert{\rm c}} + v_{\rm i}}\, \mu({\rm c}) + 3\, \frac{W_{+{\rm i}\vert{\rm i}}}{W_{-{\rm i}\vert{\rm i}} + v_{\rm i}}\, \mu({\rm i}) \, , \label{mu_i-A}
\eea
which satisfy the normalization condition~(\ref{norm_tip_prob}).  Now, the conditional probabilities~(\ref{M_cond_prob}) can be obtained as
\bea
\mu({\rm c}\vert{\rm c}) &=& \frac{W_{+{\rm c}\vert{\rm c}}}{W_{-{\rm c}\vert{\rm c}} + v_{\rm c}}\, , \label{mu_cc-A}\\
\mu({\rm c}\vert{\rm i}) &=& \frac{W_{+{\rm i}\vert{\rm c}}}{W_{-{\rm i}\vert{\rm c}} + v_{\rm i}}\, \frac{\mu({\rm c})}{\mu({\rm i})} \, , \label{mu_ci-A}\\
\mu({\rm i}\vert{\rm c}) &=& \frac{W_{+{\rm c}\vert{\rm i}}}{W_{-{\rm c}\vert{\rm i}} + v_{\rm c}}\, \frac{\mu({\rm i})}{\mu({\rm c})} \, , \label{mu_ic-A}\\
 \mu({\rm i}\vert{\rm i})&=&\frac{W_{+{\rm i}\vert{\rm i}}}{W_{-{\rm i}\vert{\rm i}} + v_{\rm i}}\, , \label{mu_ii-A}
\eea
which satisfy the normalization conditions
\be
\mu({\rm c}\vert p) + 3\, \mu({\rm i}\vert p) = 1 \qquad\mbox{for} \quad p={\rm c} \ \ \mbox{and}\ \ p={\rm i} \, .
\ee
The mean growth velocity is thus given by Eq.~(\ref{M_velocity}) in terms of the partial velocities (\ref{v_c-A})-(\ref{v_i-A}) and the tip probabilities (\ref{mu_c-A})-(\ref{mu_i-A}).  The bulk probabilities of finding the nucleotides in the bulk of the chain are then computed with Eqs.~(\ref{M_bulk_prob})  using the conditional probabilities (\ref{mu_cc-A})-(\ref{mu_ii-A}) or, equivalently, with
\bea
\bar\mu({\rm c}) =\frac{v_{\rm c}}{v}\, \mu({\rm c}) \qquad\mbox{and}\qquad \bar\mu({\rm i}) =\frac{v_{\rm i}}{v}\, \mu({\rm i})
\label{bulk_prob-A}
\eea
in terms of the tip probabilities~(\ref{mu_c-A})-(\ref{mu_i-A}), the partial velocities (\ref{v_c-A})-(\ref{v_i-A}), and the mean velocity~(\ref{M_velocity}).  The bulk probabilities satisfy the normalization condition~(\ref{norm_bulk_prob}).

\subsection{Thermodynamics}

In the regime of steady growth, the entropy production~(\ref{entr-A}) is here given by
\be
\frac{1}{R}\frac{d_{\rm i}S}{dt} = \sum_{p\, p'}
\left[W_{+p\vert p'} \, \mu(p') - W_{-p\vert p'} \, \mu(p'\vert p) \, \mu(p) \right]
 \ln\frac{W_{+p\vert p'} \, \mu(p')}{W_{-p\vert p'} \, \mu(p'\vert p) \, \mu(p)} \geq 0
\label{M_entr-A}
\ee
with $p,p'\in\{{\rm c},{\rm i},{\rm i},{\rm i}\}$, in terms of the transition rates~(\ref{rates-A}) given by Eqs.~(\ref{Wp+})-(\ref{denom}) and the probabilities~(\ref{mu_c-A})-(\ref{mu_ii-A}).  This expression is equivalent to Eq.~(\ref{entr-prod}) with the mean growth velocity~(\ref{M_velocity}), the free-energy driving force~(\ref{eps-M}) and the conditional Shannon disorder per nucleotide~(\ref{D-M}) \cite{GA14}.

\subsection{Equilibrium}

For exonuclease-deficient polymerases, the partial velocities as well as the mean velocity are vanishing at the thermodynamic equilibrium: $v_{\rm c}=v_{\rm i}= v=0$.  Accordingly, Eqs.~(\ref{mu_c-A})-(\ref{mu_i-A}) for the tip probabilities form a closed homogeneous set of equations, which admits a non-zero solution under the condition
\be
\left\vert
\begin{array}{cc}
z_{{\rm c}\vert{\rm c}}-1 & 3\, z_{{\rm c}\vert{\rm i}} \\
z_{{\rm i}\vert{\rm c}} & 3\, z_{{\rm i}\vert{\rm i}}-1 
\end{array}
\right\vert = 0 \, ,
\label{M_eq}
\ee
where
\be
z_{p\vert p'} \equiv \frac{W_{+p\vert p'}}{W_{-p\vert p'}} = \frac{K_{\rm P}\, [{\rm dNTP}] \, Q_p}{K_{p\vert p'} \, [{\rm P}]\, Q_{p'}}
\label{z-W}
\ee
for $p,p'=$ c or i.  For a given pyrophosphate concentration [P], the equilibrium condition~(\ref{M_eq}) selects a critical value $[{\rm dNTP}]_{\rm eq}$ for the nucleotide concentration.  This critical concentration can be written as in Eq.~(\ref{dNTP-eq-M}) in terms of the new variable $\delta$.  Substituting this expression into Eq.~(\ref{M_eq}) and solving for $\delta$ if $K_{{\rm c}\vert{\rm c}}\ll K_{{\rm i}\vert{\rm c}},K_{{\rm i}\vert{\rm i}}$, we obtain $\delta \simeq -3 \, K_{{\rm c}\vert{\rm c}}^2/(K_{{\rm c}\vert{\rm i}}K_{{\rm i}\vert{\rm c}})$, hence the critical concentration~(\ref{dNTP-eq-M}).

To get the error probability at equilibrium, we should notice that, if the equilibrium tip probabilities are solutions of
\be
\sum_{p'} z_{p\vert p'} \, \mu_{\rm eq}(p') = \mu_{\rm eq}(p) \, ,
\ee
the bulk probabilities satisfy
\be
\sum_{p} \frac{\bar\mu_{\rm eq}(p)}{\mu_{\rm eq}(p)} \, z_{p\vert p'} =\frac{\bar\mu_{\rm eq}(p')}{\mu_{\rm eq}(p')}\, ,
\ee
for $p,p'\in\{{\rm c},{\rm i},{\rm i},{\rm i}\}$. Solving these equations, we find that the error probability, which is defined by Eq.~(\ref{M_error}) in terms of the bulk probabilities, is given by
\be
\eta_{\rm eq,M} = \frac{z_{{\rm c}\vert{\rm c}}-1}{z_{{\rm c}\vert{\rm c}}+3 \, z_{{\rm i}\vert{\rm i}}-2} \, .
\label{M_error-A}
\ee
Using Eq.~(\ref{z-W}), we have that
\be
z_{{\rm c}\vert{\rm c}} = 1+\delta \gg z_{{\rm i}\vert{\rm i}} = \frac{K_{{\rm c}\vert{\rm c}}}{K_{{\rm i}\vert{\rm i}}}  (1+\delta) \, .
\ee
Since $\vert\delta\vert\ll 1$, we finally obtain that the error probability~(\ref{M_error-A}) is evaluated as $\eta_{\rm eq,M}\simeq - \delta$, hence Eq.~(\ref{eta-eq-M}).

\subsection{Full speed regime}

In the full speed regime, the detachment rates become negligible with respect to the attachment rates, $W_{-p\vert p'}=0$,  so that Eqs.~(\ref{v_c-A})-(\ref{v_i-A}) directly give the partial velocities as
\bea
v_{\rm c} &=& W_{+{\rm c}\vert{\rm c}} + 3\, W_{+{\rm i}\vert{\rm c}} \, , \label{v_c-FS}\\
v_{\rm i} &=& W_{+{\rm c}\vert{\rm i}} + 3\, W_{+{\rm i}\vert{\rm i}} \, . \label{v_i-FS}
\eea
On the other hand, Eqs.~(\ref{mu_c-A})-(\ref{mu_i-A}) for the tip probabilities can be solved to get
\bea
\mu({\rm c}) &=& \frac{W_{+{\rm c}\vert{\rm i}}}{W_{+{\rm c}\vert{\rm i}}+3\, W_{+{\rm i}\vert{\rm c}}} \, , \label{mu_c-FS}\\
\mu({\rm i}) &=& \frac{W_{+{\rm i}\vert{\rm c}}}{W_{+{\rm c}\vert{\rm i}}+3\, W_{+{\rm i}\vert{\rm c}}} \, , \label{mu_i-FS}
\eea
which satisfy the normalization condition~(\ref{norm_tip_prob}).

Since moreover the attachment rate of a correct base pair after the incorporation of a correct base pair is typically larger than the other ones, we have that $v_{\rm c}\gg v_{\rm i}$ and $\mu({\rm c})\gg \mu({\rm i})$ and the mean velocity can be approximated by $v\simeq v_{\rm c}\mu({\rm c})$.  Using Eqs.~(\ref{v_c-FS}) and~(\ref{mu_c-FS}) with $W_{+{\rm c}\vert{\rm c}} \gg 3\, W_{+{\rm i}\vert{\rm c}}$, the mean growth velocity is approximated by
\be
v_{\infty,{\rm M}} \simeq \frac{W_{+{\rm c}\vert{\rm c}}\, W_{+{\rm c}\vert{\rm i}}}{W_{+{\rm c}\vert{\rm i}}+3\, W_{+{\rm i}\vert{\rm c}}} \, .
\ee
Furthermore supposing $W_{+{\rm c}\vert{\rm i}} \gg 3\, W_{+{\rm i}\vert{\rm c}}$, which is equivalent to taking $\mu({\rm c})\simeq 1$, we find Eq.~(\ref{v-infty-M0}).

Now, the error probability is defined by Eq.~(\ref{M_error}), which is combined with Eq.~(\ref{bulk_prob-A}) to obtain
\be
\eta_{\infty,{\rm M}} = 3 \, \bar\mu({\rm i}) = 3 \, \mu({\rm i}) \, \frac{v_{\rm i}}{v} \simeq 3 \, \frac{\mu({\rm i}) \, v_{\rm i}}{\mu({\rm c}) \, v_{\rm c}} \, .
\ee
Substituting Eqs.~(\ref{v_c-FS})-(\ref{v_i-FS}) and Eqs.~(\ref{mu_c-FS})-(\ref{mu_i-FS}), we get an expression for the error probability in terms of the attachment rates.  Again since the attachment rate $W_{+{\rm c}\vert{\rm c}}$ is typically larger than the other ones, the error probability can be evaluated by
\be
\eta_{\infty,{\rm M}} \simeq 3 \, \frac{W_{+{\rm i}\vert{\rm c}}}{W_{+{\rm c}\vert{\rm c}}}\, \left( 1 + 3\, \frac{W_{+{\rm i}\vert{\rm i}}}{W_{+{\rm c}\vert{\rm i}}}\right) \, .
\ee
Replacing with the expressions~(\ref{M_W_c})-(\ref{M_W_i}) for the attachment rates, we finally obtain the error probability~(\ref{eta-infty-M}) in the full speed regime.

\subsection{Back to the Bernoulli-chain model}

In the case where the rates no longer depend on the previously incorporated nucleotide, i.e.,
\bea
&& W_{\pm{\rm c}\vert{\rm c}}=W_{\pm{\rm c}\vert{\rm i}}\equiv W_{\pm{\rm c}} \, , \\
&& W_{\pm{\rm i}\vert{\rm c}}=W_{\pm{\rm i}\vert{\rm i}}\equiv W_{\pm{\rm i}} \, , 
\eea
the partial velocities are equal to the mean velocity, $v_{\rm c}=v_{\rm i}= v$, and the probabilities satisfy
\bea
&& \mu({\rm c}\vert{\rm c}) = \mu({\rm c}\vert{\rm i}) = \mu({\rm c}) = \bar\mu({\rm c}) \, , \\
&& \mu({\rm i}\vert{\rm c}) = \mu({\rm i}\vert{\rm i}) = \mu({\rm i}) = \bar\mu({\rm i}) \, , 
\eea
so that we recover all the results of the Bernoulli-chain model if the approximations are otherwise comparable.

\section{The algorithm for simulating DNA replication}
\label{AppC}

The stochastic process of DNA replication is simulated at the single-molecule level with Gillespie's algorithm~\cite{G76,G77}.

Prior to the simulation, a long enough random sequence $\alpha=n_1n_2\cdots n_L$ is generated for the template.  For given nucleotide concentrations, the attachment and detachment rates~(\ref{Wp+})-(\ref{Wp-}) are calculated for every possible events.  There are five possible transitions that may happen to the copy $\omega=m_1m_2\cdots m_l$: the attachment of four possible nucleotides $m_{l+1}\in\{{\rm A},{\rm C},{\rm G},{\rm T}\}$ or the detachment of the ultimate nucleotide $m_l$ of the copy.  The rates~(\ref{Wp+})-(\ref{Wp-}) depend on the previously incorporated nucleotide, as well as on at most three consecutive nucleotides $n_{l-1}n_ln_{l+1}$ of the template.

Accordingly, at each step of the process, the length $l$ of the copy being known, the nucleotides 
\be
{m_{l-1}m_l \qquad\,\atop n_{l-1}\, n_l \, n_{l+1}}
\ee
conditioning the next event are determined.  The random time interval $\Delta t$ until the next event is exponentially distributed according to
\be
p(\Delta t) = \Gamma_l \, \exp(-\Gamma_l \, \Delta t) \, ,
\ee
with
\be
\Gamma_l = \sum_{m_{l+1}} W^{\rm p}_{+m_{l+1} m_l\atop \ \, n_{l+1}\, n_l} + W^{\rm p}_{\quad\, -m_l m_{l-1}\atop n_{l+1}\, n_l \, n_{l-1}} \, .
\ee
This random time interval is thus obtained as
\be
\Delta t = -\frac{1}{\Gamma_l}\, \ln x 
\ee
with a uniformly distributed random variable $x\in[0,1]$.  Another independent such random variable $y\in[0,1]$ is used to determine the transition occurring among the five possible ones according to the branching probabilities:
\bea
P_{+m_{l+1}} &=& \frac{1}{\Gamma_l}\, W^{\rm p}_{+m_{l+1} m_l\atop \ \, n_{l+1}\, n_l} \qquad\mbox{with}\quad m_{l+1}\in\{{\rm A},{\rm C},{\rm G},{\rm T}\} \, , \\
P_{-m_l} &=& \frac{1}{\Gamma_l}\, W^{\rm p}_{\quad\, -m_l m_{l-1}\atop n_{l+1}\, n_l \, n_{l-1}} \, .
\eea
The change of the free-energy driving force is given by Eqs.~(\ref{eps})-(\ref{eps_l}).

The procedure is repeated for many successive steps to obtain a long enough copy sequence $\omega=m_1m_2\cdots m_L$.  

The different properties of interest are computed by statistics over a large enough sample of so-generated copy sequences.

\vskip 2 cm

\end{widetext}


\end{document}